\documentclass[10pt,journal,compsoc]{IEEEtran}

\usepackage[T1]{fontenc}
\usepackage[utf8]{inputenc}
\usepackage{amsmath,amsfonts,amssymb}
\usepackage{array}
\usepackage{booktabs}
\usepackage{graphicx}
\usepackage{multirow}
\usepackage{url}
\usepackage[table]{xcolor}
\usepackage{makecell}
\usepackage{capt-of}

\usepackage{tikz}
\usetikzlibrary{positioning, shapes.geometric, arrows.meta}
\usepackage{colortbl}
\usepackage{arydshln}
\usepackage{wrapfig}
\usepackage{pdfpages}

\makeatletter
\newcommand{\thickhline}{%
    \noalign{\ifnum0=`}\fi\hrule height 1.0pt
    \futurelet\reserved@a\@xhline
}
\makeatother

\newcommand{\pub}[1]{{\color{gray}{\fontsize{4.5pt}{5.0pt}\selectfont [#1]}}}

\definecolor{FPrivacy}{HTML}{4C78A8}

\definecolor{FUnlearn}{HTML}{59A14F}

\definecolor{FGen}{HTML}{F28E2B}

\definecolor{FCaptcha}{HTML}{B07AA1}

\definecolor{FProvenance}{HTML}{E15759}

\colorlet{myheader}{gray!30}
\colorlet{myrowcolor}{gray!10}

\definecolor{capbox}{HTML}{F8E6D0}
\definecolor{rootgray}{HTML}{E6E6E6}

\newcommand{\method}[1]{\emph{#1}}

\title{Adversarial Attacks for Good: A Survey of Proactive Protection across the Visual Content Lifecycle}
\author{Jiaming Zhang, Boyang Chen, Zherui Li, Fuyao Zhang, Xinyu Yan, Hong Xi Tae, Wenwen He, Xuan Wang, Siqi Guo, Junhao Dong, Kun Wang, Hanxun Huang, Yige Li, Xingjun Ma, Yang Cao, Lingjuan Lyu, and Wei Yang Bryan Lim%
\IEEEcompsocitemizethanks{
\IEEEcompsocthanksitem J. Zhang, B. Chen, X. Yan, Z. Li, F. Zhang, H. X. Tae, W. He, X. Wang, S. Guo, J. Dong, K. Wang and W. Y. B. Lim are with the College of Computing and Data Science,
Nanyang Technological University, Singapore. W. Y. B. Lim is the corresponding author.
\IEEEcompsocthanksitem H. Huang is with the School of Computing and Information Systems,
University of Melbourne, Australia.
\IEEEcompsocthanksitem Y. Li is with the School of Computing and Information Systems,
Singapore Management University, Singapore.
\IEEEcompsocthanksitem X. Ma is with the Institute of Trustworthy Embodied AI,
Fudan University, China.
\IEEEcompsocthanksitem Y. Cao is with the Department of Computer Science,
School of Computing, Institute of Science Tokyo, Japan.
\IEEEcompsocthanksitem L. Lyu is with Sony AI, Japan.
}}

\begin{document}

\includepdf[pages=1, pagecommand={\thispagestyle{empty}}]{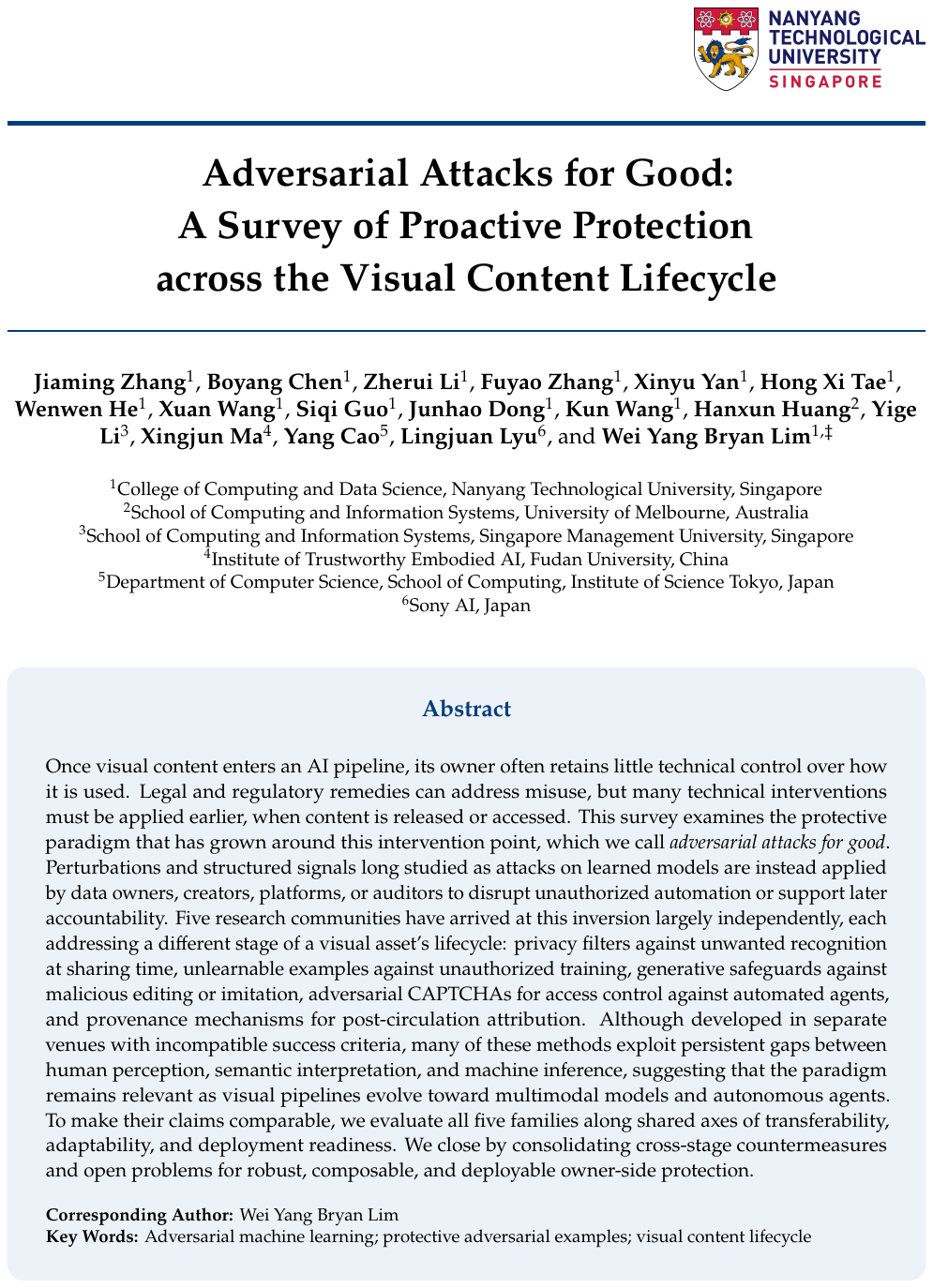}
\setcounter{page}{1}

\IEEEtitleabstractindextext{%
\begin{abstract}
Once visual content enters an AI pipeline, its owner often retains little technical control over how it is used. Legal and regulatory remedies can address misuse, but many technical interventions must be applied earlier, when content is released or accessed. This survey examines the protective paradigm that has grown around this intervention point, which we call \emph{adversarial attacks for good}. Perturbations and structured signals long studied as attacks on learned models are instead applied by data owners, creators, platforms, or auditors to disrupt unauthorized automation or support later accountability. Five research communities have arrived at this inversion largely independently, each addressing a different stage of a visual asset's lifecycle: privacy filters against unwanted recognition at sharing time, unlearnable examples against unauthorized training, generative safeguards against malicious editing or imitation, adversarial CAPTCHAs for access control against automated agents, and provenance mechanisms for post-circulation attribution. Although developed in separate venues with incompatible success criteria, many of these methods exploit persistent gaps between human perception, semantic interpretation, and machine inference, suggesting that the paradigm remains relevant as visual pipelines evolve toward multimodal models and autonomous agents. To make their claims comparable, we evaluate all five families along shared axes of transferability, adaptability, and deployment readiness. Across the lifecycle, we find that most protections are still validated mainly against static or weakly adaptive adversaries, while evidence beyond controlled benchmarks remains scarce. We close by consolidating cross-stage countermeasures and open problems for robust, composable, and deployable owner-side protection.
\end{abstract}
}

\maketitle
\IEEEdisplaynontitleabstractindextext
\IEEEpeerreviewmaketitle

\section{Introduction}\label{sec:introduction}

The rapid proliferation of multimodal foundation models, generative pipelines, and autonomous AI agents has made the automated processing of visual content a basic feature of digital infrastructure. These systems operate at a scale and with capabilities that have produced a structural asymmetry: content creators and rights holders control their visual assets until those assets enter an AI pipeline, but have no practical authority over what the pipeline does with them afterward. The legal system has begun to register this condition as litigation. Creators and publishers have brought multiple cases, several still pending, alleging that their visual work was scraped, reproduced, or repurposed for model training without authorization, and major regulatory frameworks now require AI developers to disclose the provenance of training data.\footnote{Representative cases include \textit{Andersen v.\ Stability AI} (N.D.\ Cal., filed 2023) and \textit{Getty Images v.\ Stability AI} (D.\ Del., filed 2023); on the regulatory side, the EU AI Act (Regulation (EU) 2024/1689) requires providers of general-purpose AI models to publish training-content summaries and honor machine-readable copyright reservations.}
These proceedings are symptoms rather than anomalies: litigation and regulation respond to misuse after it occurs, whereas the intervention points available to a content owner lie before the pipeline, not inside it.

One such intervention point takes its name from a foundational discovery in machine learning. Szegedy et al.~\cite{szegedy2013intriguing} demonstrated that perturbations imperceptible to humans can cause deep neural networks to misclassify inputs with high confidence. These perturbations were named adversarial examples and subsequently studied for over a decade as a security threat, an attack surface to be measured and defended against. That framing, however, presupposes who is imposing the perturbation and why. If the party applying the signal is not an adversary seeking misclassification but a data owner seeking to prevent unauthorized automated use, the objective inverts: induced model failure becomes the protection goal rather than the attack objective. This inversion is what we call \emph{adversarial attacks for good}, and a growing body of work has explored this direction~\cite{asnani2026proactive, al2024adversarial}. Adversarial examples expose a structural blind spot of gradient-trained models, one that is present in every such model, from early image classifiers to the multimodal systems and autonomous agents now at the center of digital infrastructure, and that has not been eliminated by scale or architectural change. At its core, adversarial attacks for good is an argument about paradigm: as long as AI pipelines rely on gradient-trained models, this blind spot persists, and the protective approach extends to successive systems as readily as the underlying vulnerability does.

\begin{figure*}[!t]
    \centering
    \includegraphics[width=\linewidth]{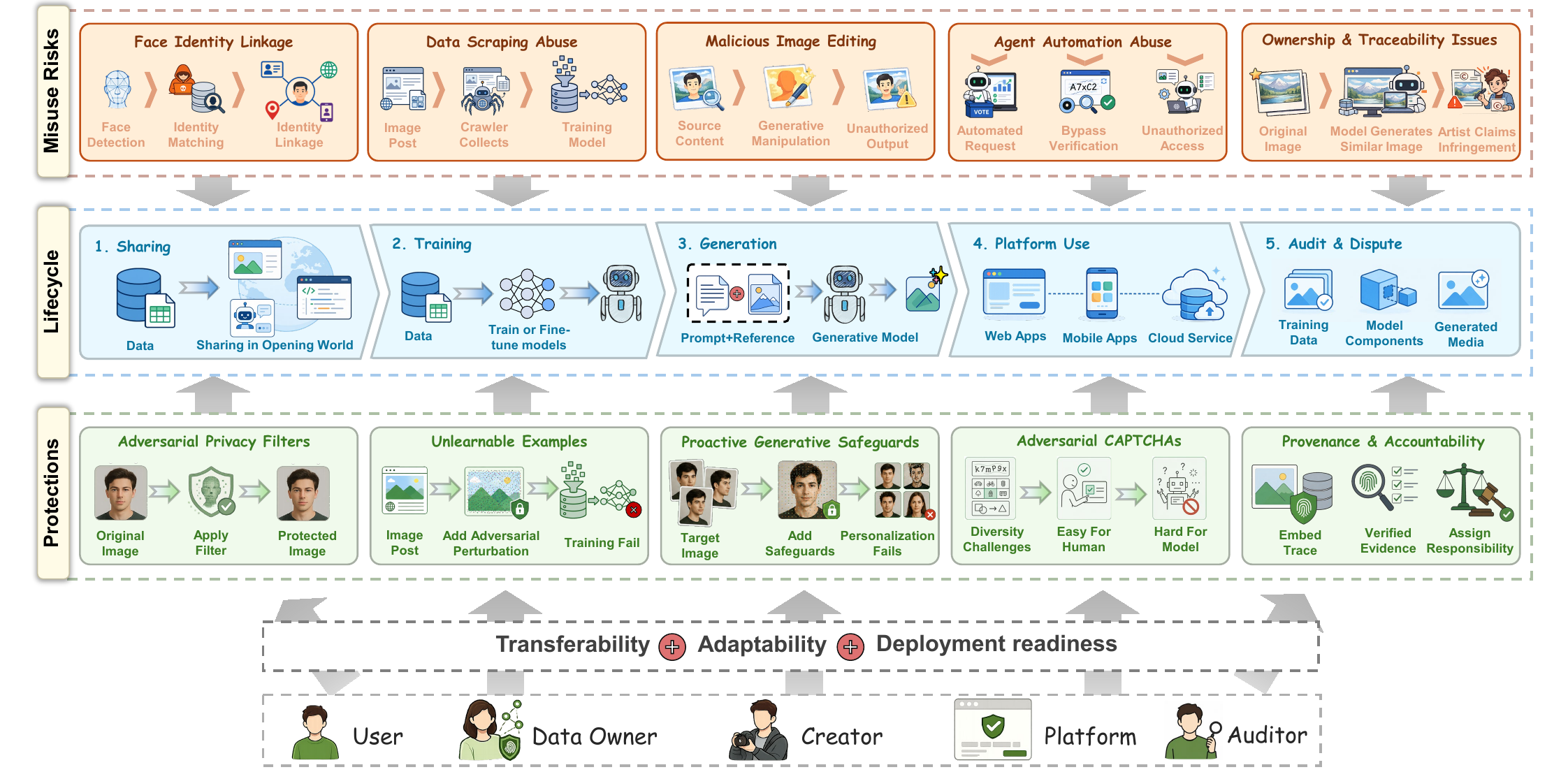}
    \caption{Overview of misuse risks, lifecycle stages, and corresponding protection mechanisms, assessed along three axes: transferability ($L_1$), adaptability ($L_2$), and deployment readiness ($L_3$).}
    \label{fig:overview}
\end{figure*}

Protection is applied before the asset enters the pipeline, but it must take effect inside the pipeline, at whichever stage the anticipated misuse occurs. The specific form the inversion takes therefore depends on the stage being defended. We identify five lifecycle stages, each pairing a distinct misuse risk with a corresponding mechanism family: adversarial privacy filters against unwanted recognition at sharing time, unlearnable examples against unauthorized training at release time, generative safeguards against manipulation and imitation by generative models, adversarial CAPTCHAs against automated abuse of online services, and provenance mechanisms for attribution after circulation. The survey is organized accordingly (Fig.~\ref{fig:overview}). Our scope is defined by two conditions: the protected asset is visual, and the protection mechanism is adversarial, operating through the structured-signal sensitivity of learned models. Differential privacy, federated learning, cryptographic access control, and research that treats adversarial perturbations as threats rather than protective tools fall outside this boundary.

Despite operating on the same mathematical property, the five communities that independently developed protective uses of adversarial mechanisms have evolved largely in isolation, publishing in separate venues, adopting incompatible vocabularies, and applying incommensurable threat models and success criteria~\cite{Laishram2025Privacy, li2025surveyunlearnabledata, deng2025surveyadvsvisual, app13074602, zhao2025sokwatermarking}. No existing survey places these families side by side (Table~\ref{tab:positioning}). This incommensurability is precisely what demands a common evaluation vocabulary; we provide one through three shared axes, transferability ($L_1$), adaptability ($L_2$), and deployment readiness ($L_3$), defined formally in Section~\ref{sec:background}.

\begin{table}[!t]
  \centering
  \scriptsize
  \setlength{\tabcolsep}{1.5pt}
  \renewcommand{\arraystretch}{1.05}
  \caption{Comparison of this survey with related surveys on coverage and analytical approach
  (\checkmark~= addressed; $\circ$~= partial; $\times$~= not addressed).}
  \label{tab:positioning}
  \begin{tabular}{@{}>{\centering\arraybackslash}p{0.45cm}>{\raggedright\arraybackslash}p{2.45cm}>{\centering\arraybackslash}p{1.00cm}>{\centering\arraybackslash}p{0.95cm}>{\centering\arraybackslash}p{1.40cm}>{\centering\arraybackslash}p{1.40cm}@{}}
    \toprule
    Ref. & Focus & \makecell{Lifecycle\\framing}
      & \makecell{Families\\(of 5)}
      & \makecell{Adaptive\\robustness} & \makecell{Deployment\\readiness} \\
    \midrule
    \multicolumn{6}{l}{\textit{Broad surveys}} \\
    \cite{asnani2026proactive}
      & proactive schemes & $\times$ & 3 & $\times$ & $\times$ \\
    \cite{al2024adversarial}
      & broad adv. ML & $\times$ & 3 & $\times$ & $\circ$ \\
    \midrule
    \multicolumn{6}{l}{\textit{Family-specific surveys}} \\
    \cite{wenger2023sok}
      & facial privacy & $\times$ & 1 & \checkmark & $\circ$ \\
    \cite{li2025surveyunlearnabledata}
      & unlearnable examples & $\times$ & 1 & \checkmark & $\circ$ \\
    \cite{nguyenle2025surveyproactive}
      & deepfake defense & $\times$ & 2 & $\circ$ & $\times$ \\
    \cite{deng2025surveyadvsvisual}
      & AIGC defense & $\times$ & 2 & $\times$ & $\times$ \\
    \cite{app13074602}
      & CAPTCHA design & $\times$ & 1 & $\circ$ & $\times$ \\
    \cite{zhao2025sokwatermarking}
      & watermarking & $\times$ & 1 & \checkmark & $\circ$ \\
    \midrule
    \rowcolor{gray!15}
      & \textbf{This survey} & \checkmark & \textbf{5} & \checkmark & \checkmark \\
    \bottomrule
  \end{tabular}
\end{table}

This survey makes three contributions.
\begin{enumerate}
\item \textbf{A durable protective paradigm.} We introduce adversarial attacks for good as a unified paradigm. Its operative principle, the perceptual gap between human observers and learned models, is a structural property of AI systems rather than an artifact of any particular architecture or training method.
\item \textbf{A lifecycle taxonomy.} We provide, to our knowledge, the first unified account of adversarial privacy filters, unlearnable examples, generative safeguards, adversarial CAPTCHAs, and provenance mechanisms as successive stages of a single protective lifecycle.
\item \textbf{A shared evaluation framework.} We compare all five families along three axes: transferability ($L_1$), adaptability ($L_2$), and deployment readiness ($L_3$), making threat models and robustness claims across the five communities directly commensurable.
\end{enumerate}

The remainder of this survey is organized as follows. Section~\ref{sec:background} formalizes protective adversarial transformations, and defines the comparison axes $L_1$--$L_3$. Sections~\ref{sec:privacy} through~\ref{sec:provenance} review the five lifecycle stages in order. Section~\ref{sec:conclusion} consolidates cross-stage countermeasures and open problems and concludes the survey.

\section{A Unified Framework for Adversarial Protection}
\label{sec:background}

\begin{figure}[t]
  \centering
  \includegraphics[width=0.70\columnwidth]{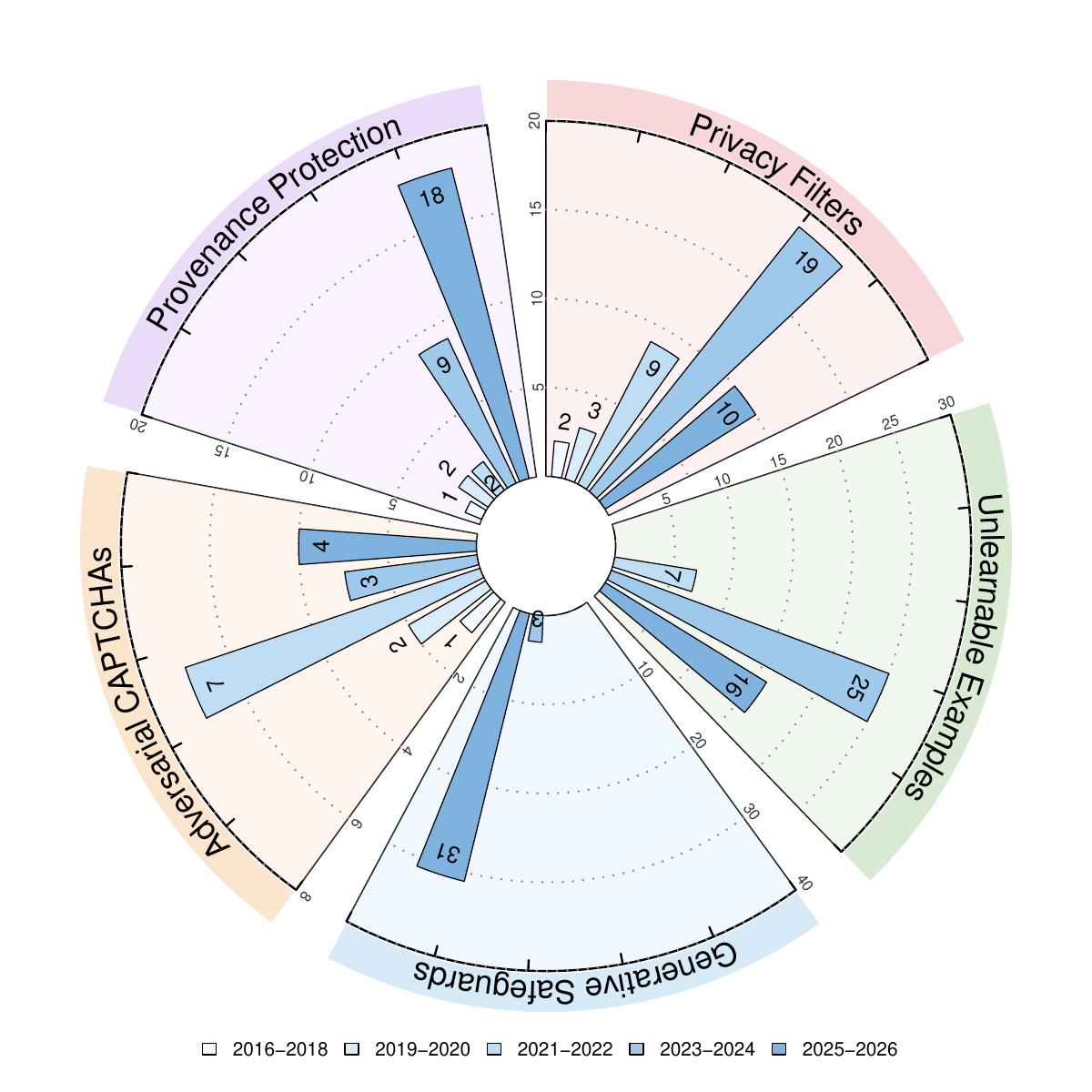}
  \caption{Publication counts of surveyed papers by family and time period.}
  \label{fig:papers_number}
\end{figure}

This section establishes the two devices used throughout the survey: the template behind all protective adversarial mechanisms (Section~\ref{subsec:template}), and the axes $L_1$--$L_3$ along which Sections~\ref{sec:privacy}--\ref{sec:provenance} compare them (Section~\ref{subsec:axes}). No familiarity with adversarial machine learning is assumed. Fig.~\ref{fig:papers_number} shows the distribution of surveyed papers by family and year.

\begin{figure*}[!t]
\centering
\begin{tikzpicture}[font=\small, scale=1.00, transform shape]

\node (f1) [draw, fill=FPrivacy!30, rounded corners, text width=3cm, align=center] at (0,18mm)
{Adversarial Privacy\\Filters\\(\S\ref{sec:privacy})};

\node (f2) [below=16mm of f1, draw, fill=FUnlearn!30, rounded corners, text width=3cm, align=center]
{Unlearnable\\Examples\\(\S\ref{sec:unlearnable})};

\node (f3) [below=16mm of f2, draw, fill=FGen!30, rounded corners, text width=3cm, align=center]
{Proactive Generative\\Safeguards\\(\S\ref{sec:generative})};

\node (f4) [below=9mm of f3, draw, fill=FCaptcha!30, rounded corners, text width=3cm, align=center]
{Adversarial\\CAPTCHAs\\(\S\ref{sec:captcha})};

\node (f5) [below=16mm of f4, draw, fill=FProvenance!30, rounded corners, text width=3cm, align=center]
{Adversarial\\Provenance and\\Accountability\\(\S\ref{sec:provenance})};

\begin{scope}[every node/.append style={font=\scriptsize}]

\node (f1l1) [right=10mm of f1, yshift=8mm, draw, fill=FPrivacy!20, rounded corners, text width=34mm, align=center] {Implicit Pixel-level};
\node (f1l2) [below=2mm of f1l1, draw, fill=FPrivacy!20, rounded corners, text width=34mm, align=center] {Explicit Semantic};
\node (f1l3) [below=2mm of f1l2, draw, fill=FPrivacy!20, rounded corners, text width=34mm, align=center] {Structured Local};

\draw[-latex] (f1.east) -- ++(4mm,0) |- (f1l1.west);
\draw[-latex] (f1.east) -- ++(4mm,0) |- (f1l2.west);
\draw[-latex] (f1.east) -- ++(4mm,0) |- (f1l3.west);

\node (f2l1) [right=10mm of f2, draw, fill=FUnlearn!20, rounded corners, text width=34mm, align=center, yshift=13mm] {Error-Based Optimization};
\node (f2l2) [below=1mm of f2l1, draw, fill=FUnlearn!20, rounded corners, text width=34mm, align=center] {Training-Guided Protection};
\node (f2l3) [below=1mm of f2l2, draw, fill=FUnlearn!20, rounded corners, text width=34mm, align=center] {Structured Shortcuts};
\node (f2l4) [below=3mm of f2l3, draw, fill=FUnlearn!20, rounded corners, text width=34mm, align=center] {Generative \mbox{Unlearnable} Examples};

\draw[-latex] (f2.east) -- ++(4mm,0) |- (f2l1.west);
\draw[-latex] (f2.east) -- ++(4mm,0) |- (f2l2.west);
\draw[-latex] (f2.east) -- ++(4mm,0) |- (f2l3.west);
\draw[-latex] (f2.east) -- ++(4mm,0) |- (f2l4.west);

\node (f3l1) [right=10mm of f3, draw, fill=FGen!20, rounded corners, text width=34mm, align=center, yshift=7mm] {Editing Immunization};
\node (f3l2) [below=3mm of f3l1, draw, fill=FGen!20, rounded corners, text width=34mm, align=center] {Subject \mbox{Personalization} Safeguards};

\draw[-latex] (f3.east) -- ++(4mm,0) |- (f3l1.west);
\draw[-latex] (f3.east) -- ++(4mm,0) |- (f3l2.west);

\node (f4l1) [right=10mm of f4, draw, fill=FCaptcha!20, rounded corners, text width=34mm, align=center, yshift=9mm] {Character-Based CAPTCHAs};
\node (f4l2) [below=2mm of f4l1, draw, fill=FCaptcha!20, rounded corners, text width=34mm, align=center] {Image-Based CAPTCHAs};
\node (f4l3) [below=1mm of f4l2, draw, fill=FCaptcha!20, rounded corners, text width=34mm, align=center] {Reasoning-Based CAPTCHAs};

\draw[-latex] (f4.east) -- ++(4mm,0) |- (f4l1.west);
\draw[-latex] (f4.east) -- ++(4mm,0) |- (f4l2.west);
\draw[-latex] (f4.east) -- ++(4mm,0) |- (f4l3.west);

\node (f5l1) [right=10mm of f5, draw, fill=FProvenance!20, rounded corners, text width=34mm, align=center, yshift=16mm] {Training Asset Tracing};
\node (f5l2) [below=2mm of f5l1, draw, fill=FProvenance!20, rounded corners, text width=34mm, align=center] {Model Ownership Verification};
\node (f5l3) [below=4mm of f5l2, draw, fill=FProvenance!20, rounded corners, text width=34mm, align=center] {Media Provenance Verification};
\node (f5l4) [below=4mm of f5l3, draw, fill=FProvenance!20, rounded corners, text width=34mm, align=center] {Evidence \mbox{Manipulation} Attacks};

\end{scope}

\draw[-latex] (f5.east) -- ++(4mm,0) |- (f5l1.west);
\draw[-latex] (f5.east) -- ++(4mm,0) |- (f5l2.west);
\draw[-latex] (f5.east) -- ++(4mm,0) |- (f5l3.west);
\draw[-latex] (f5.east) -- ++(4mm,0) |- (f5l4.west);

\node (p11) [right=6mm of f1l1, draw, fill=FPrivacy!10, rounded corners, text width=9cm, align=left, inner sep=2pt, font=\tiny\linespread{0.75}\selectfont]
{
PPVR-AT~\cite{wu2018towards};
Fawkes~\cite{shan2020fawkes};
APF~\cite{zhang2020adversarial};
LowKey~\cite{cherepanovalowkey};
SocialGuard~\cite{xue2021socialguard};
Low-Mid AP~\cite{zhang2023low};
ADAF~\cite{wu2023towards};
CamPro~\cite{zhu2024campro};
IFPC-GA~\cite{liu2024enhancing};
BFR-Obfuscation~\cite{zhang2024transferable};
VIP~\cite{meftah2025vip};
ReasonBreak~\cite{zhang2025disrupting};
GeoShield~\cite{liu2026geoshield}
};

\node (p12) [right=6mm of f1l2, draw, fill=FPrivacy!10, rounded corners, text width=9cm, align=left, inner sep=2pt, font=\tiny\linespread{0.75}\selectfont]
{
DeID-GAN~\cite{kuang2021effective};
Adv-Makeup~\cite{yin2021adv};
CLIP2Protect~\cite{shamshad2023clip2protect};
DiffProtect~\cite{liu2023diffprotect};
AdvFace~\cite{wang2023privacy};
3D-Adv Makeup~\cite{lyu20233d};
3D-Aware DeID~\cite{cao2023achieving};
GIFT~\cite{li2024transferable};
Adv-Diffusion~\cite{liu2024adv};
DiffAM~\cite{sun2024diffam};
Diff-Privacy~\cite{he2024diff};
MG-FPP~\cite{shamshad2024makeup};
StyleAdv~\cite{le2024styleadv};
SD4Privacy~\cite{an2024sd4privacy};
Adv3D-Diffusion~\cite{yang2025adversarial};
AdvCloak~\cite{liu2025advcloak};
CRFD~\cite{zhou2025crfd};
Machine Pareidolia~\cite{le2026machine}
};

\node (p13) [right=6mm of f1l3, draw, fill=FPrivacy!10, rounded corners, text width=9cm, align=left, inner sep=2pt, font=\tiny\linespread{0.75}\selectfont]
{
Accessorize~\cite{sharif2016accessorize};
Adv-Face De-ID~\cite{chatzikyriakidis2019adversarial};
TIP-IM~\cite{yang2021towards};
AFR T-Shirt~\cite{lyko2021adversarial};
IdentityMask~\cite{wen2022identitymask};
OPOM~\cite{zhong2022opom};
AMT-GAN~\cite{hu2022protecting};
Collaborative FPP~\cite{pan2023collaborative};
Diversity Mask~\cite{chow2024diversity};
AMK~\cite{ying2025reversible};
ErasableMask~\cite{shen2025erasablemask};
DualTAP~\cite{zhang2025dualtap}
};

\node (p21) [
right=6mm of f2l1,
draw,
fill=FUnlearn!10,
rounded corners,
text width=9cm,
align=left,
inner sep=2pt,
font=\tiny\linespread{0.75}\selectfont
]
{
AP~\cite{fowl2021adversarialexamples};
EM~\cite{huang2021unlearnableexamples};
ULEO-GrayAugs~\cite{liu2021goinggrayscale};
REM~\cite{fu2022robustunlearnable};
TSM-UE~\cite{fang2024rethinking};
SEM~\cite{liu2024stableunlearnable};
ARMOR~\cite{gong2026armorshielding};
ALP~\cite{liu2023securingbiomedical}
};

\node (p22) [
right=6mm of f2l2,
draw,
fill=FUnlearn!10,
rounded corners,
text width=9cm,
align=left,
inner sep=2pt,
font=\tiny\linespread{0.75}\selectfont
]
{
SEP~\cite{chen2023selfensembleprotection};
NTGA~\cite{yuan2021neuraltangent};
EntF~\cite{wen2023adversarialtraining};
DH~\cite{meng2024semanticdeep};
MI-UE~\cite{zhu2026whydo};
TUE~\cite{ren2023transferableunlearnable};
AUE/AAP~\cite{wang2024efficientavailability};
UC~\cite{zhang2023unlearnableclusters};
14A~\cite{chen2024oneforall};
MEM~\cite{liu2024multimodalunlearnable};
UnSeg~\cite{sun2024unsegone};
T2UE~\cite{ma2025t2uegenerating};
VTG~\cite{li2025versatiletransferable};
GUE~\cite{liu2024gametheoretic};
BAIT~\cite{li2026whenpriors};
PUE~\cite{wang2025provablyunlearnable};
Segue~\cite{zhang2025segueside}
};

\node (p23) [
right=6mm of f2l3,
draw,
fill=FUnlearn!10,
rounded corners,
text width=9cm,
align=left,
inner sep=2pt,
font=\tiny\linespread{0.75}\selectfont
]
{
LSP~\cite{yu2022availabilityattacks};
AR~\cite{sandovalsegura2022autoregressive};
OPS~\cite{wu2023onepixelshortcut};
CUDA~\cite{sadasivan2023cudaconvolution};
IRP~\cite{huang2024leveragingimperfect};
KBS~\cite{li2025kspace};
UMed~\cite{lin2024safeguardingmedical};
TUE-V~\cite{wu2025temporalunlearnable}
};

\node (p24) [
right=6mm of f2l4,
draw,
fill=FUnlearn!10,
rounded corners,
text width=9cm,
align=left,
inner sep=2pt,
font=\tiny\linespread{0.75}\selectfont
]
{
AdvDM~\cite{liang2023adversarial};
Glaze~\cite{shan2023glaze};
Nightshade~\cite{shan2024nightshade};
DiffProtect-UE~\cite{xue2024toward};
StyleGuard~\cite{li2025styleguard};
StyleProtect~\cite{tang2025styleprotect};
FastProtect~\cite{ahn2025nearly};
Anti-DreamBooth~\cite{van2023anti};
InMark~\cite{liu2024counteringpersonalized};
MetaCloak~\cite{liu2024metacloak};
HAAD~\cite{xu2025h};
GoodAC~\cite{xu2025harnessing};
DisDiff~\cite{liu2024disrupting};
ACE~\cite{zheng2023targeted};
AntiPure~\cite{yang2025towards}
};

\node (p31) [
right=6mm of f3l1,
draw,
fill=FGen!10,
rounded corners,
text width=9cm,
align=left,
inner sep=2pt,
font=\tiny\linespread{0.75}\selectfont
]
{
PhotoGuard~\cite{salman2023raising};
PCA~\cite{guo2025gray};
DiffusionGuard~\cite{choi2025diffusionguard};
DiffVax~\cite{ozden2026diffvax};
Distraction~\cite{lo2024distraction};
EditShield~\cite{chen2024editshield};
DCT-Shield~\cite{bala2025dct};
DeContext~\cite{shen2025decontext};
DIA~\cite{hong2025dia};
Edit Away~\cite{wang2025edit};
Pixel Is Not a Barrier~\cite{shih2025pixel};
PSFD~\cite{zeng2025psfd};
Semantic Mismatch~\cite{dong2025semantic};
TarPro~\cite{shen2026tarpro};
Universal Image Immunization~\cite{lee2026universal};
BlurGuard~\cite{kim2026blurguard};
Transferable Defense~\cite{zhang2026towards};
AdvPaint~\cite{jeon2025advpaint};
Anti-Inpainting~\cite{guo2025anti};
PromptFlare~\cite{na2025promptflare};
My Face Is Mine, Not Yours~\cite{yam2025my};
Beauty and the Beast~\cite{huang2026beauty};
Cascading Pathway Disruption~\cite{wang2026safeguarding};
I2VGuard~\cite{gui2025i2vguard};
Anti-I2V~\cite{vu2026anti};
Vid-Freeze~\cite{chowdhury2025vid};
DORMANT~\cite{zhou2025dormant};
Silence Is Golden~\cite{gan2025silence};
SyncBreaker~\cite{zhang2026syncbreaker}
};

\node (p32) [
right=6mm of f3l2,
draw,
fill=FGen!10,
rounded corners,
text width=9cm,
align=left,
inner sep=2pt,
font=\tiny\linespread{0.75}\selectfont
]
{
IDProtector~\cite{song2025idprotector};
DLADiff~\cite{jia2025dladiff};
AIR~\cite{lyu2025transferable};
Targeted Ensemble Defense~\cite{hu2025targeted};
Adapter Shield~\cite{jia2026adapter};
};

\node (p41) [
right=6mm of f4l1,
draw,
fill=FCaptcha!10,
rounded corners,
text width=9cm,
align=left,
inner sep=2pt,
font=\tiny\linespread{0.75}\selectfont
]
{
rCAPTCHA~\cite{zhang2020robust};
Spatial Smoothing~\cite{matsuura2021adversarial};
MBAGP~\cite{dankwa2021securing};
AECAPTCHA~\cite{wang2021text};
aCAPTCHA~\cite{shi2022adversarial};
RTC~\cite{shao2022robust};
ACG~\cite{11288041}
};

\node (p42) [
right=6mm of f4l2,
draw,
fill=FCaptcha!10,
rounded corners,
text width=9cm,
align=left,
inner sep=2pt,
font=\tiny\linespread{0.75}\selectfont
]
{
DeepCAPTCHA~\cite{osadchy2017no};
Robust CAPTCHA Generator~\cite{ardhita2020robust};
Capture-the-bot~\cite{hitaj2020capture};
Diff-CAPTCHA~\cite{jiang2023diff};
DAC~\cite{du2025defensive}
};

\node (p43) [
right=6mm of f4l3,
draw,
fill=FCaptcha!10,
rounded corners,
text width=9cm,
align=left,
inner sep=2pt,
font=\tiny\linespread{0.75}\selectfont
]
{
TICS~\cite{jia2022tics};
zxCAPTCHA~\cite{trong2023new,dinh2023zxcaptcha};
IllusionCAPTCHA~\cite{ding2025illusioncaptcha};
Next-Gen CAPTCHAs~\cite{liu2026next}
};

\node (p51) [
right=6mm of f5l1,
draw,
fill=FProvenance!10,
rounded corners,
text width=9cm,
align=left,
inner sep=2pt,
font=\tiny\linespread{0.75}\selectfont
]
{
Radioactive Data~\cite{sablayrolles2020radioactivedata};
Dataset Inference~\cite{maini2021datasetinference};
DVBW~\cite{li2023blackbox};
SSCL-BW~\cite{wang2025ssclbwsample};
X-Mark~\cite{kulkarni2026xmarksaliency}
};

\node (p52) [
right=6mm of f5l2,
draw,
fill=FProvenance!10,
rounded corners,
text width=9cm,
align=left,
inner sep=2pt,
font=\tiny\linespread{0.75}\selectfont
]
{
DNN Watermark~\cite{zhang2018protectingintellectual};
IPN Watermark~\cite{zhang2020modelwatermarking};
Deep Watermarking~\cite{zhang2021deepmodel};
Wide-Flat GAN WM~\cite{fei2023wideflat};
Free Fine-tuning WM~\cite{wang2023freefine};
CycleGAN Watermark~\cite{lin2024cycleganwatermarking};
PlugMark~\cite{chen2025plugmarkplug};
VLA-Mark~\cite{liu2025vlamarkcross};
SWAP~\cite{yang2025swapcopyrightauditing};
Cert-LAS~\cite{qi2026certlascertified};
VLA Backdoor Ownership~\cite{sun2026backdoorbasedownership};
LoRA-Key~\cite{wang2026lorakeyuser};
SIF~\cite{zhao2026sifsemanticallydistribution}
};

\node (p53) [
right=6mm of f5l3,
draw,
fill=FProvenance!10,
rounded corners,
text width=9cm,
align=left,
inner sep=2pt,
font=\tiny\linespread{0.75}\selectfont
]
{
ROBIN~\cite{huang2024robinrobust};
ConceptWM~\cite{lei2024watermarkingvisual};
Traceable Adv. Examples~\cite{li2024dualprotection};
Invisible Adv. WM~\cite{wang2024invisibleadversarial};
ZoDiac~\cite{zhang2024attackresilient};
NoisePrints~\cite{goren2025noiseprintsdistortionfree};
Video Signature~\cite{huang2025videosignatureimplicit};
BitMark~\cite{kerner2025bitmarkwatermarkingbitwise};
Info-Theoretic AIGI Detector~\cite{zhang2025adversariallyrobustai};
CSGuard~\cite{lai2026csguardforgeryresistant};
RWP~\cite{liu2026rwprobust};
AEON~\cite{muneer2026aeonadaptive};
Dual-Guard~\cite{xie2026dualguard};
Robust Content WM~\cite{zhu2026robustcontentwatermarking}
};

\node (p54) [
right=6mm of f5l4,
draw,
fill=FProvenance!10,
rounded corners,
text width=9cm,
align=left,
inner sep=2pt,
font=\tiny\linespread{0.75}\selectfont
]
{
Watermark Radioactivity Eval.~\cite{dubiski2025arewatermarks};
Dataset Copyright Evasion~\cite{gao2025datasetcopyrightevasion};
FT-Traceability Benchmark~\cite{wang2025evaluatingdatasetwatermarking};
MarkSweep~\cite{cao2026marksweepnobox};
Forensic-Stealth WM Removal~\cite{goonatilake2026removingwatermarkis};
Fragile Reconstruction~\cite{jiang2026fragilereconstructionadversarial};
RAVEN~\cite{shamshad2026ravenerasinginvisible};
Frequency-Domain WM Attack~\cite{wang2026breakingwatermarksfrequency}
};

\draw[-latex] (f1l1.east) -- ++(4mm,0) |- (p11.west);
\draw[-latex] (f1l2.east) -- ++(4mm,0) |- (p12.west);
\draw[-latex] (f1l3.east) -- ++(4mm,0) |- (p13.west);

\draw[-latex] (f2l1.east) -- ++(4mm,0) |- (p21.west);
\draw[-latex] (f2l2.east) -- ++(4mm,0) |- (p22.west);
\draw[-latex] (f2l3.east) -- ++(4mm,0) |- (p23.west);
\draw[-latex] (f2l4.east) -- ++(4mm,0) |- (p24.west);

\draw[-latex] (f3l1.east) -- ++(4mm,0) |- (p31.west);
\draw[-latex] (f3l2.east) -- ++(4mm,0) |- (p32.west);

\draw[-latex] (f4l1.east) -- ++(4mm,0) |- (p41.west);
\draw[-latex] (f4l2.east) -- ++(4mm,0) |- (p42.west);
\draw[-latex] (f4l3.east) -- ++(4mm,0) |- (p43.west);

\draw[-latex] (f5l1.east) -- ++(4mm,0) |- (p51.west);
\draw[-latex] (f5l2.east) -- ++(4mm,0) |- (p52.west);
\draw[-latex] (f5l3.east) -- ++(4mm,0) |- (p53.west);
\draw[-latex] (f5l4.east) -- ++(4mm,0) |- (p54.west);

\end{tikzpicture}
\caption{Three-level taxonomy of protective adversarial mechanisms, organized by family, subcategory, and representative works.}
\label{fig:taxonomy}
\end{figure*}
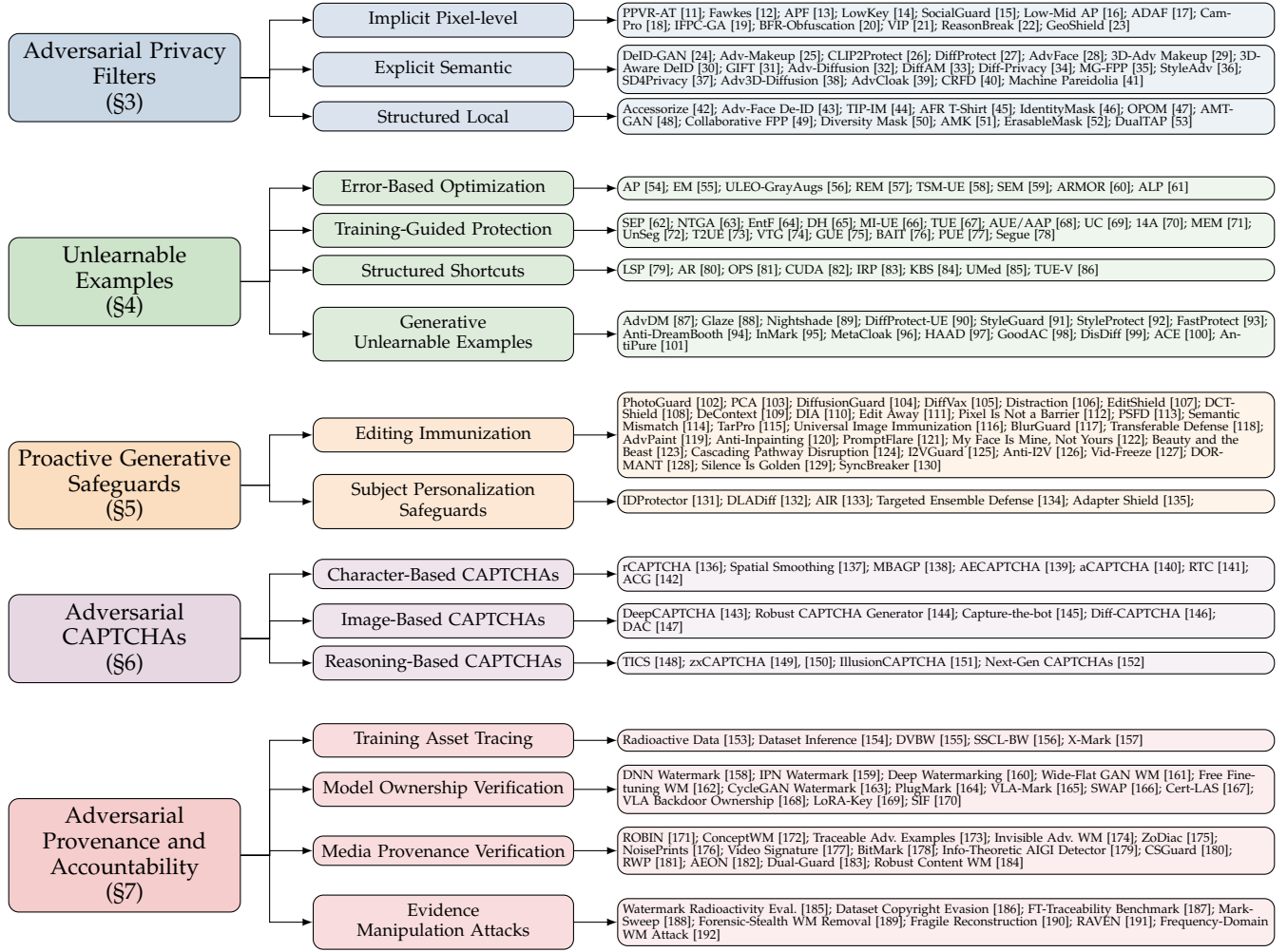

\subsection{Protective Adversarial Transformations}
\label{subsec:template}
Szegedy et al.~\cite{szegedy2013intriguing} discovered that deep networks can be steered by signals people cannot see: for almost any input $x$, there is a perturbation $\delta$ so faint that $x+\delta$ looks identical to $x$, yet the model's output flips, $f(x+\delta) \neq f(x)$. Two facts about these \emph{adversarial examples} matter here. They are structural, not incidental: gradient-trained models rely on faint input statistics that human perception discards, so a decade of architectural change has not removed them~\cite{al2024adversarial,asnani2026proactive}. And they \emph{transfer}: a perturbation computed on one model often deceives another trained independently~\cite{papernot2016transferability}, so the effect does not require access to its eventual target.

\noindent
\hspace*{2em} Attack research treats these properties as a threat; protection reverses the roles, and the reversal is easiest to see by following a single portrait photo after it is posted online (Fig.~\ref{fig:overview}). A recognition service may index the face and link it to an identity. A scraper may fold the photo into a training corpus.
A generative model may re-synthesize the person in scenes that never happened. The bots doing the scraping must first pass the hosting platform's human-verification challenges. And once copies circulate, the owner may need to prove where the image came from or how it was used. Sections~\ref{sec:privacy}--\ref{sec:provenance} survey the protections built for these five moments, and they all share one move: embedding an adversarial signal in the asset \emph{before} release.

\hspace*{2em} Throughout, we call the asset owner the \emph{protector} and the operator of the unwanted pipeline the \emph{adversary}, inverting the classical usage.
The protector controls the asset only until release; the adversary controls everything afterward, including manipulations $g \in \mathcal{G}$ (re-encoding, restoration, deliberate countermeasures) applied before the pipeline runs. The protector therefore applies a transformation $T$ before release,
\begin{equation}
\tilde{x} = T(x),
\qquad
\underbrace{d\big(x, \tilde{x}\big) \le \epsilon}_{\text{utility}},
\qquad
\underbrace{F\big(g(\tilde{x})\big) \neq y}_{\text{protection}},
\label{eq:template}
\end{equation}
where $F \in \mathcal{F}$ is the unwanted pipeline and $y$ the output it is built to obtain: the utility condition keeps the asset valuable to its human audience, under a small budget $\epsilon$, while the protection condition makes the pipeline fail regardless.
Each family reads $x$, $F$, $d$, and failure in its own terms; each section's opening states its own reading. The one deliberate departure is provenance (Section~\ref{sec:provenance}): applied when misuse can no longer be prevented, it replaces failure with verification, $V(g(\tilde{x})) = 1$ for a designated verifier $V$, turning the signal into evidence rather than sabotage.

\subsection{A Common Evaluation Lens: $L_1$--$L_3$}
\label{subsec:axes}

The five communities report robustness in mutually incompatible vocabularies; the axes make their claims commensurable, recording what a method assumes about the pipeline $\mathcal{F}$ ($L_1$), which manipulations $\mathcal{G}$ it has been validated under ($L_2$), and how mature its evidence is ($L_3$).

\textbf{$L_1$: transferability} grades the access under which protection is built and verified: \emph{white-box} (the deployed target itself), \emph{gray-box} (shared components such as a public backbone), or \emph{black-box} (surrogates only, so that surrogate-to-target transfer carries the entire claim). The polarity is inverted relative to attacks: there, white-box is the informative worst case; here, black-box is the demanding level, because protectors can rarely inspect the service that will process their asset. A face cloak, for instance, must defeat the commercial API actually deployed, not the surrogate it was optimized against. In the accountability stage, $L_1$ instead records the verification access available to the party proving misuse.

\textbf{$L_2$: adaptability} grades the manipulations $g$ a method has been shown to survive: \emph{static} validation against a fixed pipeline, \emph{routine} media operations applied without knowledge of the protection (compression, resizing, re-encoding), or informed \emph{adaptive} countermeasures that target it (purification, adapted training, signal removal or forgery). Because the protector commits the signal at release and cannot re-optimize it afterward, robustness against uninformed adversaries says little about informed ones; this is where protection claims most often collapse.

\textbf{$L_3$: deployment readiness} grades the maturity of the evidence: \emph{laboratory} demonstrations on in-house models, \emph{external} validation against systems the authors do not control (commercial APIs, online services, human users), or sustained \emph{operational} use. The axis separates what is feasible from what is genuinely available.

Sections~\ref{sec:privacy} through~\ref{sec:provenance} instantiate these axes under their own threat models in \emph{domain-specific vocabulary}.

\section{Adversarial Privacy Filters}
\label{sec:privacy}
\begin{table*}[!htbp]
\centering
\caption{Taxonomy of adversarial privacy filter methods for privacy protection.}
\label{tab:privacy_filter_taxonomy}
\vspace{-5pt}
\tiny
\setlength{\tabcolsep}{11pt}
\renewcommand{\arraystretch}{1.0}

\begin{tabular}{l|cll|c}
\thickhline
\rowcolor{myheader}

\textbf{Paper}
& \textbf{$L_1$-Transferability}
& \multicolumn{1}{c}{\textbf{$L_2$-Adaptability}}
& \multicolumn{1}{c}{\textbf{$L_3$-Deployment Readiness}}
& \textbf{Target Model} \\

\hline

\multicolumn{5}{l}{\textbf{Implicit Pixel-level}} \\
\hdashline

\rowcolor{myrowcolor}
\method{PPVR-AT}\pub{ECCV'18}~\cite{wu2018towards}
& White-box & Non-Interactive
& Client-side Pre-upload
& Traditional FR \\

\method{Fawkes}\pub{USENIX'20}~\cite{shan2020fawkes}
& Black-box & Non-Interactive
& Client-side Pre-upload
& Traditional FR \\

\rowcolor{myrowcolor}
\method{APF}\pub{ACM MM'20}~\cite{zhang2020adversarial}
& Black-box & Non-Interactive
& Client-side Pre-upload
& Traditional FR \\

\method{LowKey}\pub{ICLR'21}~\cite{cherepanovalowkey}
& Black-box & Non-Interactive
& Client-side Pre-upload
& Traditional FR \\

\rowcolor{myrowcolor}
\method{SocialGuard}\pub{JISA'21}~\cite{xue2021socialguard}
& Black-box & Non-Interactive
& Client-side Pre-upload
& Traditional FR \\

\rowcolor{myrowcolor}
\method{Low-Mid AP}\pub{IS'23}~\cite{zhang2023low}
& Black-box & Non-Interactive
& Client-side Pre-upload
& Traditional FR \\

\method{ADAF}\pub{Arxiv'23}~\cite{wu2023towards}
& Black-box & Content-Adaptive
& Client-side Pre-upload
& Traditional FR \\

\rowcolor{myrowcolor}
\method{CamPro}\pub{NDSS'24}~\cite{zhu2024campro}
& Black-box & Non-Interactive
& Physical Device-side
& Traditional FR \\

\method{IFPC-GA}\pub{ICASSP'24}~\cite{liu2024enhancing}
& Black-box & Non-Interactive
& Client-side Pre-upload
& Traditional FR \\

\rowcolor{myrowcolor}
\method{BFR-Obfuscation}\pub{ICML'24}~\cite{zhang2024transferable}
& Black-box & Non-Interactive
& Client-side Pre-upload
& Traditional FR \\

\method{VIP}\pub{Arxiv'25}~\cite{meftah2025vip}
& White-box & Content-Adaptive
& Client-side Pre-upload
& MLLM \\

\rowcolor{myrowcolor}
\method{ReasonBreak}\pub{ICLR'26}~\cite{zhang2025disrupting}
& Black-box & Content-Adaptive
& Client-side Pre-upload
& MLLM \\

\method{GeoShield}\pub{AAAI'26}~\cite{liu2026geoshield}
& Black-box & Content-Adaptive
& Client-side Pre-upload
& MLLM \\

\hline

\multicolumn{5}{l}{\textbf{Explicit Semantic}} \\
\hdashline

\rowcolor{myrowcolor}
\method{DeID-GAN}\pub{ACM MM'21}~\cite{kuang2021effective}
& Black-box & Non-Interactive
& Client-side Pre-upload
& Traditional FR \\

\method{Adv-Makeup}\pub{IJCAI'21}~\cite{yin2021adv}
& Black-box & Non-Interactive
& Client-side Pre-upload
& Traditional FR \\

\rowcolor{myrowcolor}
\method{CLIP2Protect}\pub{CVPR'23}~\cite{shamshad2023clip2protect}
& Black-box & Non-Interactive
& Client-side Pre-upload
& Traditional FR \\

\method{DiffProtect}\pub{PR'26}~\cite{liu2023diffprotect}
& Black-box & Non-Interactive
& Client-side Pre-upload
& Traditional FR \\

\rowcolor{myrowcolor}
\method{AdvFace}\pub{CVPR'23}~\cite{wang2023privacy}
& White-box & Non-Interactive
& Platform / Cloud-side
& Traditional FR \\

\method{3D-Adv Makeup}\pub{TPAMI'23}~\cite{lyu20233d}
& Black-box & Non-Interactive
& Client-side Pre-upload
& Traditional FR \\

\rowcolor{myrowcolor}
\method{3D-Aware DeID}\pub{MM Asia'23}~\cite{cao2023achieving}
& Black-box & Non-Interactive
& Client-side Pre-upload
& Traditional FR \\

\method{GIFT}\pub{ACM MM'24}~\cite{li2024transferable}
& Black-box & Non-Interactive
& Client-side Pre-upload
& Traditional FR \\

\rowcolor{myrowcolor}
\method{Adv-Diffusion}\pub{AAAI'24}~\cite{liu2024adv}
& Black-box & Non-Interactive
& Client-side Pre-upload
& Traditional FR \\

\method{DiffAM}\pub{CVPR'24}~\cite{sun2024diffam}
& Black-box & Non-Interactive
& Client-side Pre-upload
& Traditional FR \\

\rowcolor{myrowcolor}
\method{Diff-Privacy}\pub{TCSVT'24}~\cite{he2024diff}
& Black-box & Reversible
& Client-side Pre-upload
& Traditional FR \\

\method{MG-FPP}\pub{ECCV'24}~\cite{shamshad2024makeup}
& Black-box & Non-Interactive
& Client-side Pre-upload
& Traditional FR \\

\rowcolor{myrowcolor}
\method{StyleAdv}\pub{PoPETs'24}~\cite{le2024styleadv}
& Black-box & Reversible
& Client-side Pre-upload
& Traditional FR \\

\method{SD4Privacy}\pub{ICME'24}~\cite{an2024sd4privacy}
& Black-box & Non-Interactive
& Client-side Pre-upload
& Traditional FR \\

\rowcolor{myrowcolor}
\method{Adv3D-Diffusion}\pub{ICIA'25}~\cite{yang2025adversarial}
& Black-box & Non-Interactive
& Client-side Pre-upload
& Traditional FR \\

\method{AdvCloak}\pub{PR'25}~\cite{liu2025advcloak}
& Black-box & Non-Interactive
& Client-side Pre-upload
& Traditional FR \\

\rowcolor{myrowcolor}
\method{CRFD}\pub{ESWA'25}~\cite{zhou2025crfd}
& Black-box & Reversible
& Client-side Pre-upload
& Traditional FR \\

\method{Machine Pareidolia}\pub{AAAI'26}~\cite{le2026machine}
& Black-box & Non-Interactive
& Client-side Pre-upload
& Traditional FR \\

\hline
\multicolumn{5}{l}{\textbf{Structured Local}} \\
\hdashline

\rowcolor{myrowcolor}
\method{Accessorize}\pub{CCS'16}~\cite{sharif2016accessorize}
& White-box & Non-Interactive
& Physical Device-side
& Traditional FR \\

\method{Adv-Face De-ID}\pub{ICIP'19}~\cite{chatzikyriakidis2019adversarial}
& White-box & Non-Interactive
& Client-side Pre-upload
& Traditional FR \\

\rowcolor{myrowcolor}
\method{TIP-IM}\pub{ICCV'21}~\cite{yang2021towards}
& Black-box & Non-Interactive
& Client-side Pre-upload
& Traditional FR \\

\method{AFR T-Shirt}\pub{ICCCI'21}~\cite{lyko2021adversarial}
& Black-box & Non-Interactive
& Physical Device-side
& Traditional FR \\

\rowcolor{myrowcolor}
\method{IdentityMask}\pub{TCSVT'22}~\cite{wen2022identitymask}
& Black-box & Reversible
& Client-side Pre-upload
& Traditional FR \\

\method{OPOM}\pub{TPAMI'22}~\cite{zhong2022opom}
& Black-box & Non-Interactive
& Client-side Pre-upload
& Traditional FR \\

\rowcolor{myrowcolor}
\method{AMT-GAN}\pub{CVPR'22}~\cite{hu2022protecting}
& Black-box & Non-Interactive
& Client-side Pre-upload
& Traditional FR \\

\method{Collaborative FPP}\pub{ICIC'23}~\cite{pan2023collaborative}
& Black-box & Non-Interactive
& Platform / Cloud-side
& Traditional FR \\

\rowcolor{myrowcolor}
\method{Diversity Mask}\pub{PETS'24}~\cite{chow2024diversity}
& Black-box & Content-Adaptive
& Client-side Pre-upload
& Traditional FR \\

\method{AMK}\pub{ACM MM'25}~\cite{ying2025reversible}
& White-box & Reversible
& Client-side Pre-upload
& MLLM \\

\rowcolor{myrowcolor}
\method{ErasableMask}\pub{TMM'25}~\cite{shen2025erasablemask}
& Black-box & Reversible
& Client-side Pre-upload
& Traditional FR \\

\method{DualTAP}\pub{ECCV'26}~\cite{zhang2025dualtap}
& Black-box & Content-Adaptive
& Platform / Cloud-side
& Visual Agent \\

\thickhline
\end{tabular}
\end{table*}

\textbf{Adversarial privacy filters} are proactive mechanisms that protect facial identity before images enter recognition pipelines, injecting either imperceptible pixel-level noise or semantically meaningful appearance changes to mislead automated inference while preserving human-perceived quality. As facial recognition (FR) systems pervade social platforms, surveillance infrastructure, and increasingly multimodal large language models (MLLMs) and visual agents, we compare these filters along the survey's three axes: \textbf{transferability} ($L_1$, generalizing from surrogates to inaccessible deployed models), \textbf{adaptability} ($L_2$, surviving content variation, platform processing, and supporting reversibility), and \textbf{deployment readiness} ($L_3$, maturing from client-side pre-upload toward source- and platform-side operation).

We organize existing methods by perturbation type, which most directly determines the threat model, visual fidelity, and deployment pathway. As summarized in Table~\ref{tab:privacy_filter_taxonomy}, this section reviews three categories: (1)~\textit{Implicit Pixel-Level Perturbation}, hiding adversarial signals in imperceptible noise; (2)~\textit{Explicit Semantic Perturbation}, protecting identity via natural appearance transformations; (3)~\textit{Structured Local Perturbation}, confining modifications to facial or wearable regions.

\subsection{Implicit Pixel-Level Perturbation}

Implicit pixel-level protection develops as one trajectory: it first secures black-box transferability so a cloak works against recognizers the user cannot access, then hardens that cloak to survive the processing pipelines of real platforms, and finally moves protection earlier in the capture pipeline and outward to new inference targets.

\noindent\textbf{Establishing black-box transferability.} The defining scenario is a user who wants to share a photo but cannot query the commercial recognizer that will index it, so protection needF transfer from a surrogate to an unseen deployed model. Early work assumed white-box access---\method{PPVR-AT}~\cite{wu2018towards} optimizes directly against the target. However, this assumption fails when the recognizer is proprietary. \method{Fawkes}~\cite{shan2020fawkes} and \method{LowKey}~\cite{cherepanovalowkey} remove it by nudging feature embeddings toward decoy identities through surrogate ensembles, and this transferable formulation is what lets \method{SocialGuard}~\cite{xue2021socialguard} succeed against live commercial recognition APIs rather than offline benchmarks.

\noindent\textbf{Surviving real-world platform processing.} Transferability alone is insufficient once the image is uploaded, because the platform itself re-encodes and restores it before any recognizer runs; the next question is therefore whether the perturbation survives this pipeline. Since JPEG recompression and blind face restoration erase raw pixel noise, \method{APF}~\cite{zhang2020adversarial}, \method{Low-Mid AP}~\cite{zhang2023low}, and \method{BFR-Obfuscation}~\cite{zhang2024transferable} relocate the signal into resilient frequency bands. \method{ADAF}~\cite{wu2023towards} and \method{IFPC-GA}~\cite{liu2024enhancing} complement this with content-adaptive noise budgets that concentrate perturbation where it survives while keeping the shared image visually clean.

\noindent\textbf{Extending to source-level and multimodal deployment.} With a cloak that both transfers and survives, the remaining frontier is where protection is applied and what it protects against. \method{CamPro}~\cite{zhu2024campro} moves protection off the client and into the camera sensor, obfuscating identity before an image is ever stored. In parallel, the target has shifted from traditional face recognition to multimodal models: \method{VIP}~\cite{meftah2025vip}, \method{ReasonBreak}~\cite{zhang2025disrupting}, and \method{GeoShield}~\cite{liu2026geoshield} disrupt the visual encoders and reasoning chains of MLLMs, extending the pre-upload filter from blocking identity matching to blocking higher-level inference such as geolocation.

\subsection{Explicit Semantic Perturbation}

Explicit semantic protection replaces imperceptible noise with natural transformations like digital makeup~\cite{yin2021adv,kuang2021effective}. Its \textbf{transferability} relies on \emph{black-box} assumptions, turning protection into a stylistic choice rather than a suspicious artifact. Its \textbf{adaptability} leverages diffusion models for user-friendly editing and reversibility for authorized recovery~\cite{sun2024diffam,liu2023diffprotect,le2024styleadv,zhou2025crfd}. For \textbf{deployment}, 3D-aware variants extend these tools to physical surveillance across varying viewpoints~\cite{lyu20233d,yang2025adversarial}.

\noindent\textbf{Establishing protection through natural appearance.} The motivating scenario is a user whose protection need not look like an attack, so that it survives the noise-stripping filters platforms apply and raises no suspicion. \method{Adv-Makeup}~\cite{yin2021adv} realizes this with transferable cosmetic overlays, while \method{DeID-GAN}~\cite{kuang2021effective} and \method{3D-Aware DeID}~\cite{cao2023achieving} anonymize identity while preserving expression. Diffusion models then make these edits controllable rather than fixed: \method{DiffAM}~\cite{sun2024diffam} and \method{DiffProtect}~\cite{liu2023diffprotect} synthesize the protective appearance end-to-end, and \method{CLIP2Protect}~\cite{shamshad2023clip2protect,shamshad2024makeup} lets users specify it by text prompt~\cite{liu2024adv}, while \method{GIFT}~\cite{li2024transferable} and \method{AdvCloak}~\cite{liu2025advcloak} further improve black-box semantic protection through transferable feature-level or cloak-style optimization. Control that in turn allows \method{AdvFace}~\cite{wang2023privacy} to explore platform-side protection, while \method{SD4Privacy}~\cite{an2024sd4privacy} applies generative editing in a pre-upload privacy-preserving setting.

\noindent\textbf{Supporting reversibility for authorized recovery.} Because a semantic edit alters the visible image rather than adding removable noise, the natural next question is whether a trusted party can recover the original---an adaptability property pixel-level cloaks rarely offer. \method{StyleAdv}~\cite{le2024styleadv} restores the protected face for authorized users holding the correct cryptographic key, and disentanglement-based \method{CRFD}~\cite{zhou2025crfd} and \method{Diff-Privacy}~\cite{he2024diff} suppress identity while preserving expression and lighting, so recovered photos remain usable for downstream tasks such as video conferencing.

\noindent\textbf{Extending to physical, cross-view surveillance.} The final stage carries semantic protection out of the well-lit single image and into physical surveillance, where the protector controls neither viewpoint nor lighting. \method{3D-Adv Makeup}~\cite{lyu20233d} binds adversarial cosmetics to the facial surface so evasion holds across camera angles, and \method{Adv3D-Diffusion}~\cite{yang2025adversarial} adds generative refinement to remove residual video artifacts. \method{Machine Pareidolia}~\cite{le2026machine} pushes this furthest, letting users actively shape how a security camera perceives them rather than merely evading detection.

\subsection{Structured Local Perturbation}

Structured local protection confines adversarial signals to bounded semantic regions~\cite{sharif2016accessorize}. Its \textbf{transferability} relies on \emph{black-box}, person-specific patterns in identity-critical areas~\cite{yang2021towards,zhong2022opom}. Its \textbf{adaptability} extends beyond plain cloaking to support reversibility, diversity, and multi-user coordination~\cite{shen2025erasablemask,wen2022identitymask,chow2024diversity}. The key distinction is \textbf{deployment}, as spatial locality maps easily onto physical wearables, platform filters, and visual agent protection~\cite{lyko2021adversarial,zhang2025dualtap}.

\noindent\textbf{Establishing protection through localized regions.} The founding scenario is physically realizable evasion: \method{Accessorize}~\cite{sharif2016accessorize} shows that a printed pair of glasses alone can fool a recognizer, establishing that a signal confined to a small semantic region suffices to control identity. \method{Adv-Face De-ID}~\cite{chatzikyriakidis2019adversarial} and \method{AMT-GAN}~\cite{hu2022protecting} bring this localization into image space, and \method{TIP-IM}~\cite{yang2021towards} and \method{OPOM}~\cite{zhong2022opom} make it transferable by learning person-specific patterns concentrated on identity-critical zones rather than a single model-tuned perturbation.

\noindent\textbf{Supporting recovery and multi-user coordination.} Once localized protection transfers, deploying it across a user community raises two adaptability demands the single-image view ignores: authorized recovery, and avoiding collapse when many users protect against the same service. \method{ErasableMask}~\cite{shen2025erasablemask} and \method{IdentityMask}~\cite{wen2022identitymask} embed a recoverable signal inside the local patch, \method{Diversity Mask}~\cite{chow2024diversity} regularizes users away from converging on a shared decoy identity, and \method{Collaborative FPP}~\cite{pan2023collaborative} coordinates perturbations server-side to reduce the community's aggregate identity leakage.

\noindent\textbf{Extending to physical and agentic deployment.} Finally, because the signal is spatially confined, it transfers naturally to deployment settings beyond the pre-upload image. \method{AFR T-Shirt}~\cite{lyko2021adversarial} scales the wearable from glasses to full-body clothing against video re-identification, while the target expands to automated visual agents: \method{AMK}~\cite{ying2025reversible} places reversible masks at patch-grid boundaries to exploit MLLM tokenization, and \method{DualTAP}~\cite{zhang2025dualtap} moves protection platform-side, intercepting image streams to suppress sensitive-attribute inference in tool-augmented pipelines.

\subsection{Discussion}

\noindent\textbf{Countermeasures.} Protection strength here is conditional on the adversary's post-release processing, and the manipulations that break these filters cluster around a single capability: restoring a near-clean image before the recognizer runs. JPEG recompression, blind face restoration, and diffusion-based purification attenuate pixel- and frequency-space cloaks, which is precisely why later pixel-level methods migrate the signal into resilient frequency bands and adaptive budgets~\cite{zhang2020adversarial,zhang2024transferable}. Explicit semantic and structured local perturbations resist naive noise removal because their signal is carried by natural appearance or a bounded region rather than additive noise, but they remain exposed to a complementary attack surface, switching the deployed recognizer backbone or detecting and filtering the protective pattern once it is characterized. Reversibility widens this surface further: a recoverable signal that a trusted party can invert is also a signal an adversary can target, so key management and trust become part of the threat model rather than an implementation detail. Across the three categories, then, no single design dominates; each trades protection, recoverability, fidelity, and cost against a different slice of the adversary's capabilities.

\noindent\textbf{Open problems and future directions.}
The key challenge for adversarial privacy filters is moving beyond fixed face-recognition settings toward more adaptive and compositional visual inference pipelines. First, evaluation remains insufficiently standardized. Existing methods are often tested under different models, datasets, and preprocessing settings, making robustness claims hard to compare. Future benchmarks should explicitly include adaptive restoration and recognizer-switching attacks and measure residual privacy leakage rather than only recognition failure. Second, the protection objective should expand beyond identity matching. Modern VLMs can infer sensitive attributes, locations, activities, and social context even when face recognition is disrupted. Privacy filters therefore need finer leakage metrics and more selective protection mechanisms that suppress sensitive evidence without destroying benign visual utility. Third, visual agents introduce a stronger threat model. Unlike passive recognizers, agents can combine visual cues with instructions, memory, tools, and external search, turning weak residual evidence into private conclusions or actions. Future work should evaluate privacy filters by downstream agent behavior, asking whether they can serve as trust-aware privacy layers rather than single-image cloaks.

\section{Unlearnable Examples}
\label{sec:unlearnable}

\begin{table*}[!htbp]
\centering
\caption{Taxonomy of unlearnable example methods for unauthorized training prevention.}
\label{tab:unlearnable}
\vspace{-8pt}
\tiny
\setlength{\tabcolsep}{10pt}
\renewcommand{\arraystretch}{1.04}

\begin{tabular}{l|cll|c}
\thickhline
\rowcolor{myheader}
\textbf{Paper}
& \textbf{$L_1$-Transferability}
& \multicolumn{1}{c}{\textbf{$L_2$-Adaptability}}
& \multicolumn{1}{c}{\textbf{$L_3$-Deployment Readiness}}
& \textbf{Target Model} \\
\hline

\multicolumn{5}{l}{\textbf{Error-Based Optimization}} \\
\hdashline

\rowcolor{myrowcolor}
\method{AP}\pub{NeurIPS'21}~\cite{fowl2021adversarialexamples}
& Black-box & Transformation Resistant
& Application Scenario
& Discriminative \\

\method{EM}\pub{ICLR'21}~\cite{huang2021unlearnableexamples}
& Black-box & Non-Adaptive
& Application Scenario
& Discriminative \\

\rowcolor{myrowcolor}
\method{ULEO-GrayAugs}\pub{arXiv'21}~\cite{liu2021goinggrayscale}
& Black-box & Transformation Resistant
& No Deployment Evidence
& Discriminative \\

\method{REM}\pub{ICLR'22}~\cite{fu2022robustunlearnable}
& Black-box & Training-Pipeline Resistant
& No Deployment Evidence
& Discriminative \\

\rowcolor{myrowcolor}
\method{TSM-UE}\pub{CVPR'24}~\cite{fang2024rethinking}
& Black-box & Training-Pipeline Resistant
& No Deployment Evidence
& Discriminative \\

\method{SEM}\pub{AAAI'24}~\cite{liu2024stableunlearnable}
& Black-box & Training-Pipeline Resistant
& Application Scenario
& Discriminative \\

\rowcolor{myrowcolor}
\method{ARMOR}\pub{TPAMI'26}~\cite{gong2026armorshielding}
& Black-box & Transformation Resistant
& Application Scenario
& Discriminative \\

\method{ALP}\pub{arXiv'23}~\cite{liu2023securingbiomedical}
& White-box & Non-Adaptive
& Application Scenario
& Discriminative \\

\hline
\multicolumn{5}{l}{\textbf{Training-Guided Protection}} \\
\hdashline

\rowcolor{myrowcolor}
\method{SEP}\pub{ICLR'23}~\cite{chen2023selfensembleprotection}
& Black-box & Non-Adaptive
& No Deployment Evidence
& Discriminative \\

\method{NTGA}\pub{ICML'21}~\cite{yuan2021neuraltangent}
& Black-box & Non-Adaptive
& No Deployment Evidence
& Discriminative \\

\rowcolor{myrowcolor}
\method{EntF}\pub{ICLR'23}~\cite{wen2023adversarialtraining}
& Black-box & Training-Pipeline Resistant
& No Deployment Evidence
& Discriminative \\

\method{DH}\pub{TIFS'24}~\cite{meng2024semanticdeep}
& Black-box & Transformation Resistant
& No Deployment Evidence
& Discriminative \\

\rowcolor{myrowcolor}
\method{MI-UE}\pub{ICLR'26}~\cite{zhu2026whydo}
& Black-box & Non-Adaptive
& No Deployment Evidence
& Discriminative \\

\method{TUE}\pub{ICLR'23}~\cite{ren2023transferableunlearnable}
& Black-box & Training-Pipeline Resistant
& No Deployment Evidence
& Discriminative \\

\rowcolor{myrowcolor}
\method{AUE/AAP}\pub{NeurIPS'24}~\cite{wang2024efficientavailability}
& Black-box & Training-Pipeline Resistant
& No Deployment Evidence
& Discriminative \\

\method{UC}\pub{CVPR'23}~\cite{zhang2023unlearnableclusters}
& Black-box & Non-Adaptive
& External Evaluation
& Discriminative \\

\rowcolor{myrowcolor}
\method{14A}\pub{ICML'24}~\cite{chen2024oneforall}
& Black-box & Non-Adaptive
& Application Scenario
& Discriminative \\

\method{MEM}\pub{ACM MM'24}~\cite{liu2024multimodalunlearnable}
& Black-box & Non-Adaptive
& Application Scenario
& Foundational \\

\rowcolor{myrowcolor}
\method{UnSeg}\pub{NeurIPS'24}~\cite{sun2024unsegone}
& Black-box & Non-Adaptive
& Application Scenario
& Discriminative \\

\method{T2UE}\pub{ACM MM'25}~\cite{ma2025t2uegenerating}
& Black-box & Non-Adaptive
& Workflow Design
& Foundational \\

\rowcolor{myrowcolor}
\method{VTG}\pub{NeurIPS'25}~\cite{li2025versatiletransferable}
& Black-box & Non-Adaptive
& Application Scenario
& Discriminative \\

\method{GUE}\pub{AAAI'24}~\cite{liu2024gametheoretic}
& Black-box & Training-Pipeline Resistant
& No Deployment Evidence
& Discriminative \\

\rowcolor{myrowcolor}
\method{BAIT}\pub{ICLR'26}~\cite{li2026whenpriors}
& Black-box & Training-Pipeline Resistant
& Application Scenario
& Foundational \\

\method{PUE}\pub{NDSS'25}~\cite{wang2025provablyunlearnable}
& White-box & Training-Pipeline Resistant
& No Deployment Evidence
& Discriminative \\

\rowcolor{myrowcolor}
\method{Segue}\pub{ICASSP'25}~\cite{zhang2025segueside}
& Black-box & Training-Pipeline Resistant
& Application Scenario
& Discriminative \\

\hline
\multicolumn{5}{l}{\textbf{Structured Shortcuts}} \\
\hdashline

\rowcolor{myrowcolor}
\method{LSP}\pub{KDD'22}~\cite{yu2022availabilityattacks}
& Black-box & Transformation Resistant
& Application Scenario
& Discriminative \\

\method{AR}\pub{NeurIPS'22}~\cite{sandovalsegura2022autoregressive}
& Black-box & Transformation Resistant
& No Deployment Evidence
& Discriminative \\

\rowcolor{myrowcolor}
\method{OPS}\pub{ICLR'23}~\cite{wu2023onepixelshortcut}
& Black-box & Non-Adaptive
& No Deployment Evidence
& Discriminative \\

\method{CUDA}\pub{CVPR'23}~\cite{sadasivan2023cudaconvolution}
& Black-box & Training-Pipeline Resistant
& No Deployment Evidence
& Discriminative \\

\rowcolor{myrowcolor}
\method{IRP}\pub{ECCV'24}~\cite{huang2024leveragingimperfect}
& Black-box & Purification Resistant
& No Deployment Evidence
& Discriminative \\

\method{KBS}\pub{ACM MM'25}~\cite{li2025kspace}
& Black-box & Transformation Resistant
& No Deployment Evidence
& Discriminative \\

\rowcolor{myrowcolor}
\method{UMed}\pub{arXiv'24}~\cite{lin2024safeguardingmedical}
& Black-box & Non-Adaptive
& Application Scenario
& Discriminative \\

\method{TUE-V}\pub{ICCV'25}~\cite{wu2025temporalunlearnable}
& Black-box & Non-Adaptive
& Application Scenario
& Discriminative \\

\hline
\multicolumn{5}{l}{\textbf{Generative Unlearnable Examples}} \\
\hdashline

\rowcolor{myrowcolor}
\method{AdvDM}\pub{ICML'23}~\cite{liang2023adversarial}
& Black-box & Non-Adaptive
& Application Scenario
& Generative \\

\method{Glaze}\pub{USENIX Security'23}~\cite{shan2023glaze}
& Black-box & Non-Adaptive
& External Evaluation
& Generative \\

\rowcolor{myrowcolor}
\method{Nightshade}\pub{IEEE S\&P'24}~\cite{shan2024nightshade}
& Black-box & Non-Adaptive
& Application Scenario
& Generative \\

\method{DiffProtect-UE}\pub{ICLR'24}~\cite{xue2024toward}
& Black-box & Non-Adaptive
& Application Scenario
& Generative \\

\rowcolor{myrowcolor}
\method{StyleGuard}\pub{NeurIPS'25}~\cite{li2025styleguard}
& Black-box & Purification Resistant
& Application Scenario
& Generative \\

\method{StyleProtect}\pub{CVPR'26}~\cite{tang2025styleprotect}
& Black-box & Non-Adaptive
& Application Scenario
& Generative \\

\rowcolor{myrowcolor}
\method{FastProtect}\pub{CVPR'25}~\cite{ahn2025nearly}
& Black-box & Non-Adaptive
& Application Scenario
& Generative \\

\method{Anti-DreamBooth}\pub{ICCV'23}~\cite{van2023anti}
& Black-box & Non-Adaptive
& Application Scenario
& Generative \\

\rowcolor{myrowcolor}
\method{InMark}\pub{CVPR'24}~\cite{liu2024counteringpersonalized}
& Black-box & Transformation Resistant
& Application Scenario
& Generative \\

\method{MetaCloak}\pub{CVPR'24}~\cite{liu2024metacloak}
& Black-box & Transformation Resistant
& External Evaluation
& Generative \\

\rowcolor{myrowcolor}
\method{HAAD}\pub{ACM MM'25}~\cite{xu2025h}
& Black-box & Non-Adaptive
& Application Scenario
& Generative \\

\method{GoodAC}\pub{CVPR'25}~\cite{xu2025harnessing}
& Black-box & Non-Adaptive
& Application Scenario
& Generative \\

\rowcolor{myrowcolor}
\method{DisDiff}\pub{ACM MM'24}~\cite{liu2024disrupting}
& Black-box & Non-Adaptive
& Application Scenario
& Generative \\

\method{ACE}\pub{ICLR'25}~\cite{zheng2023targeted}
& Black-box & Non-Adaptive
& Application Scenario
& Generative \\

\rowcolor{myrowcolor}
\method{AntiPure}\pub{ICCV'25}~\cite{yang2025towards}
& Black-box & Purification Resistant
& Application Scenario
& Generative \\

\thickhline
\end{tabular}
\end{table*}

\textbf{Unlearnable examples (UE)}, also studied under unlearnable data, availability protection, or data availability attacks, aim to restrict unauthorized use of released data for model training while preserving human-perceived or authorized utility. Whereas adversarial privacy filters protect facial identity before images enter recognition or identity-linkage pipelines, UE addresses the subsequent unauthorized-training stage: released images may still be scraped into downstream training pipelines, so protection shifts from blocking recognition to limiting what models can learn from the data. In discriminative settings, protected data should cause trained models to generalize poorly to clean test data; in generative settings, protection appears as degraded generation quality, failed personalization, or misdirected fine-tuning.

Table~\ref{tab:unlearnable} summarizes the surveyed techniques by \textbf{transferability}, \textbf{adaptability}, \textbf{deployment readiness}, and target model families.

The development of UE can be organized into four stages. (1) \emph{Error-based optimization} establishes the basic data-side perturbation paradigm. (2) \emph{Training-guided protection} incorporates training dynamics, representations, objectives, and model priors into UE generation. (3) \emph{Structured shortcuts} constructs non-semantic signals that models learn in place of task-relevant visual features. (4) \emph{Generative UEs} extend protection to text-to-image fine-tuning, personalization, customization, and style mimicry.

\subsection{Error-Based Optimization}

Error-based optimization grows out of availability poisoning and reframes it as personal or sensitive data protection. Its application scenario is data release before downstream training: images are modified so that later unauthorized training becomes unreliable. Its \textbf{transferability} is usually evaluated through black-box transfer to unseen downstream models, its \textbf{adaptability} pressure mainly comes from transformation resistance against preprocessing or augmentation and training-pipeline resistance against adversarial training~\cite{liu2021goinggrayscale,gong2026armorshielding,fu2022robustunlearnable,fang2024rethinking,liu2024stableunlearnable}, and its \textbf{deployment readiness} develops through extensions from personal-image protection toward sensitive-domain application scenarios such as biomedical-image protection~\cite{huang2021unlearnableexamples,liu2023securingbiomedical}.

\noindent\textbf{From poisoning to personal-data protection.}
\method{AP}~\cite{fowl2021adversarialexamples} establishes the availability-attack starting point by showing that adversarial examples can become strong training-time poisons. \method{EM}~\cite{huang2021unlearnableexamples} turns this idea toward personal image protection, where imperceptible error-minimizing noise prevents released photos from supporting unauthorized model training. The key shift is from attacking a learner to protecting a data asset before it enters the training pipeline.

\noindent\textbf{Robustness pressure.}
Once personal-data protection is established, the main question is whether the protection survives realistic training operations. \method{ULEO-GrayAugs}~\cite{liu2021goinggrayscale} shows that early UE can be weakened by grayscale filtering and augmentation, motivating more robust error-based protection. Several error-based methods instead focus on adversarial-training robustness~\cite{fu2022robustunlearnable,fang2024rethinking,liu2024stableunlearnable}, whereas \method{ARMOR}~\cite{gong2026armorshielding} emphasizes robustness to data augmentation. These works move the setting from clean benchmark training toward practical preprocessing and training pipelines.

\noindent\textbf{Practical-domain extension.}
\method{ALP}~\cite{liu2023securingbiomedical} extends input-optimization UE to biomedical images, where unauthorized training is tied to data governance, institutional reuse, and sensitive-domain release. Rather than simply adding another dataset, this application evaluates error-based UE in a sensitive-domain release scenario where unauthorized training raises privacy, compliance, and data-governance concerns.

\subsection{Training-Guided Protection}

Training-guided protection shifts UE from directly optimizing input loss to modeling what unauthorized training will learn. Its \textbf{transferability} is mainly evaluated through black-box transfer across downstream models, while later work broadens the target settings beyond standard classification and architecture-level transfer, including label-agnostic use and zero-contact protection workflows~\cite{ren2023transferableunlearnable,zhang2023unlearnableclusters,ma2025t2uegenerating}. Its \textbf{adaptability} is mainly captured by training-pipeline resistance, including learning-paradigm changes, adversarial training, pretrained backbones, and parameter recovery~\cite{ren2023transferableunlearnable,wang2024efficientavailability,liu2024gametheoretic,li2026whenpriors,wang2025provablyunlearnable}. Its \textbf{deployment readiness} develops through label-agnostic evaluation, zero-contact protection workflows, and pipeline-aware face-image release scenarios~\cite{zhang2023unlearnableclusters,ma2025t2uegenerating,zhang2025segueside}.

\noindent\textbf{Training guidance across representations, settings, and objectives.}
Early training-guided methods use the training process itself as a protection signal: \method{SEP}~\cite{chen2023selfensembleprotection} uses checkpoint ensembles, and \method{NTGA}~\cite{yuan2021neuraltangent} uses generalization approximations to guide protected-data generation. Representation-oriented work then protects data by limiting the reusable features that unauthorized models can extract from protected examples~\cite{wen2023adversarialtraining,meng2024semanticdeep,zhu2026whydo}. The target settings also broaden from standard classification to cross-paradigm learning~\cite{ren2023transferableunlearnable,wang2024efficientavailability}, label-agnostic training~\cite{zhang2023unlearnableclusters}, concept-level training~\cite{chen2024oneforall}, multimodal contrastive learning~\cite{liu2024multimodalunlearnable}, segmentation~\cite{sun2024unsegone}, and cross-domain or cross-task transfer~\cite{li2025versatiletransferable}.

\noindent\textbf{Adapting to stronger training pipelines.}
As downstream training pipelines become stronger, UE must address training choices that can weaken protection, such as switching learning paradigms, adversarial training, pretrained backbones, or parameter recovery. \method{GUE}~\cite{liu2024gametheoretic} models the protector and learner as a game to improve robustness under adversarial training. \method{BAIT}~\cite{li2026whenpriors} addresses pretrained-backbone bypass, where strong prior representations reduce the effect of traditional UE. \method{PUE}~\cite{wang2025provablyunlearnable} studies learnability under parameter recovery or uncertainty, clarifying when protected data remain costly or unreliable to learn.

\noindent\textbf{Deployment-aware workflows.}
Deployment-oriented work connects training guidance to more realistic data-release workflows. \method{UC}~\cite{zhang2023unlearnableclusters} provides label-agnostic external evaluation, addressing cases where the protector cannot assume the unauthorized learner's label taxonomy. \method{T2UE}~\cite{ma2025t2uegenerating} supports zero-contact protection workflows by generating UE from text descriptions rather than requiring access to private images. \method{Segue}~\cite{zhang2025segueside} studies pipeline-aware face-image release under side information and adversarial training, with transmission distortion considered in the release pipeline.

\subsection{Structured Shortcuts}

In deep learning, shortcut learning describes the tendency of models to rely on simple but non-causal cues rather than task-relevant features~\cite{geirhos2020shortcut}. Structured shortcuts adapt this idea to UE by explicitly constructing non-semantic signals that unauthorized models learn more readily than task-relevant visual features. The application goal is not to stop training from converging, but to make the learned rule depend on a shortcut that fails under clean deployment. Its \textbf{transferability} is mainly assessed through black-box transfer across downstream models, its \textbf{adaptability} centers on purification resistance against restoration or noise removal and transformation resistance against filtering or signal-domain processing~\cite{huang2024leveragingimperfect,li2025kspace}, and its \textbf{deployment readiness} develops as shortcut construction moves beyond image-level classification into task-structured settings such as medical segmentation and video tracking~\cite{lin2024safeguardingmedical,wu2025temporalunlearnable}.

\noindent\textbf{Constructing explicit shortcut signals.}
\method{LSP}~\cite{yu2022availabilityattacks} shows that availability attacks can create learnable shortcuts without detailed victim-model knowledge, and \method{AR}~\cite{sandovalsegura2022autoregressive} improves stability through autoregressive local structures. \method{OPS}~\cite{wu2023onepixelshortcut} shows that shortcut signals can be extremely local, even at the one-pixel level, while \method{CUDA}~\cite{sadasivan2023cudaconvolution} constructs class-wise convolutional transformations. Together, these works make shortcut learning an explicit protection mechanism rather than a byproduct of perturbation optimization.

\noindent\textbf{Robust shortcuts under purification and transformations.}
After shortcuts become explicit, the main adaptability question is whether they remain learnable after countermeasures weaken the signal. \method{IRP}~\cite{huang2024leveragingimperfect} exploits imperfections in restoration countermeasures, while \method{KBS}~\cite{li2025kspace} designs frequency-domain shortcuts to resist filtering and visible detail loss. Together, these methods treat restoration and frequency processing as expected parts of the unauthorized training path.

\noindent\textbf{Toward deployment in task-structured settings.}
\method{UMed}~\cite{lin2024safeguardingmedical} adapts shortcut design to medical segmentation, where protection must preserve task-relevant spatial structure. \method{TUE-V}~\cite{wu2025temporalunlearnable} extends shortcut protection to video object tracking, where temporal matching must be considered in protection design. These applications move structured shortcuts beyond image classification benchmarks toward downstream tasks where protection must account for spatial regions, contours, textures, or temporal correspondences.

\subsection{Generative UEs}

Generative UEs protect images that may be exploited as reference or training data in unauthorized generative pipelines. The central scenarios are style mimicry, concept learning, subject personalization, and customization services. Their \textbf{transferability} is usually black-box because the service, backbone, prompt, and customization method are outside the data owner's control. Their \textbf{adaptability} is mainly shaped by transformation resistance to compression or service-side processing and purification resistance against diffusion-based purification~\cite{liu2024metacloak,li2025styleguard,yang2025towards}. \textbf{Deployment readiness} is especially visible in artist-facing, privacy-facing, and service-facing studies~\cite{shan2023glaze,liu2024metacloak}. Generative UE methods then develop along two deployment directions: artist-facing style and concept protection and privacy-facing subject and identity protection.

\noindent\textbf{Toward artist-facing style and concept protection.}
\method{AdvDM}~\cite{liang2023adversarial} adapts adversarial perturbations to disrupt diffusion-based imitation, establishing an early path from perturbative UE to generative misuse. \method{Glaze}~\cite{shan2023glaze} turns this path into artist-facing style cloaking, while \method{Nightshade}~\cite{shan2024nightshade} extends protection from style mimicry to prompt-specific concept-text misalignment. Later style-protection methods strengthen creative-asset protection against diffusion mimicry and style extraction~\cite{xue2024toward,tang2025styleprotect}.

\noindent\textbf{Toward privacy-facing subject and identity protection.}
A parallel deployment direction protects personal subjects and identities in personalization services. \method{Anti-DreamBooth}~\cite{van2023anti} disrupts subject-driven personalization, while \method{InMark}~\cite{liu2024counteringpersonalized} targets personalized text-to-image workflows. Subsequent work broadens this privacy-facing direction to protecting personal subjects and identities across few-shot personalization, customization, and low-cost service settings~\cite{xu2025h,liu2024disrupting,zheng2023targeted,xu2025harnessing,ahn2025nearly}. This branch is therefore privacy-facing, with personalization and customization services serving as the main deployment setting rather than artist-facing style mimicry.

\noindent\textbf{Countermeasure-aware adaptation across deployment pipelines.}
Generative UE must also survive the transformations that platforms or downstream users apply before customization. \method{StyleGuard}~\cite{li2025styleguard} incorporates purification-aware style protection, \method{MetaCloak}~\cite{liu2024metacloak} improves subject-driven protection under input transformations and online service evaluation, and \method{AntiPure}~\cite{yang2025towards} directly models diffusion purification followed by customization. This adaptation pressure cuts across both artist-facing and privacy-facing deployments: protected images must remain disruptive after purification, compression, and service-side processing.

\subsection{Discussion}

\noindent\textbf{Countermeasures.}
Countermeasure studies show that UE effectiveness depends on adversary capabilities and deployment paths. Some countermeasures recover learnability before training through compression, projection, or variational autoencoder purification~\cite{liu2023imageshortcutsqueezing,geiping2023whatcan,yu2024purifyunlearnable}. Others bypass protection during training by changing objectives, such as prompt learning with cross-modal alignment~\cite{wang2025sup3}, or by detecting and filtering UE perturbations~\cite{zhu2024detectiondefense}. For generative UE, media transformations, purification-customization pipelines, resilience evaluation, and service-leakage purification further show that perturbations may be weakened before or during personalization~\cite{cao2023impressevaluating,wang2024bridgepurelimited}. Together, these countermeasures motivate cost-of-learning metrics for UE evaluation: beyond whether unauthorized learning is completely prevented, evaluation should measure the additional data, adaptation, compute, or query costs required to recover usable models.

\noindent\textbf{Open problems and future directions.}
Emerging reuse scenarios further challenge UE by shifting protection from released samples to reusable task experience. In GUI-agent customization, workflow screenshots and action traces may be reused to fine-tune agents, exposing interface states, proprietary procedures, and action policies. UE could limit such unauthorized policy learning while preserving authorized customization. In embodied intelligence, shared robot demonstrations or egocentric observations may reveal manipulation routines and spatial layouts. UE could protect such videos so that collaborators can inspect task execution while making the same data harder to reuse for imitation learning or policy training. These scenarios motivate two directions for future UE research. First, emerging reuse targets workflows, policies, and skills rather than isolated samples, calling for dataset-level, concept-level, or capability-level UE through coordinated data-side interventions. Second, controlled learnability could make UE more practical by moving beyond blanket disruption toward conditional authorization, recovery, model-conditional learnability, target steering, and provenance-aware accountability~\cite{peng2022learnabilitylock,ye2024ungeneralizable,lee2025targeteddata,wang2026reversibleunlearnable}.

\section{Proactive Generative Safeguards}
\label{sec:generative}

\begin{table*}[!htbp]
\centering
\caption{Taxonomy of proactive generative safeguards against visual misuse.}
\label{tab:generative_safeguards}
\vspace{-5pt}
\tiny
\setlength{\tabcolsep}{4.2pt}
\renewcommand{\arraystretch}{1.0}

\begin{tabular}{l|cll|c}
\thickhline
\rowcolor{myheader}
\textbf{Paper}
& \textbf{$L_1$-Transferability}
& \multicolumn{1}{c}{\textbf{$L_2$-Adaptability}}
& \multicolumn{1}{c}{\textbf{$L_3$-Deployment Readiness}}
& \textbf{Target Model} \\
\hline

\multicolumn{5}{l}{\textbf{Editing Immunization}} \\
\hdashline

\rowcolor{myrowcolor}
\method{PhotoGuard}\pub{ICML'23}~\cite{salman2023raising}
& White-box & Static
& Laboratory
& Image Editor \\

\method{PCA}\pub{TIFS'25}~\cite{guo2025gray}
& Gray-box & Adaptive
& Laboratory
& Image Editor \\

\rowcolor{myrowcolor}
\method{DiffusionGuard}\pub{ICLR'25}~\cite{choi2025diffusionguard}
& Gray-box & Adaptive
& External
& Image Editor \\

\method{DiffVax}\pub{ICLR'26}~\cite{ozden2026diffvax}
& Gray-box & Adaptive
& External
& Image Editor \\

\rowcolor{myrowcolor}
\method{Distraction}\pub{CVPR'24}~\cite{lo2024distraction}
& Gray-box & Static
& Laboratory
& Image Editor \\

\method{EditShield}\pub{ECCV'24}~\cite{chen2024editshield}
& White-box & Routine
& External
& Image Editor \\

\rowcolor{myrowcolor}
\method{DCT-Shield}\pub{ICCV'25}~\cite{bala2025dct}
& Gray-box & Adaptive
& External
& Image Editor \\

\method{DeContext}\pub{arXiv'25}~\cite{shen2025decontext}
& White-box & Static
& External
& Image Editor \\

\rowcolor{myrowcolor}
\method{DIA}\pub{ICCV'25}~\cite{hong2025dia}
& Gray-box & Adaptive
& Laboratory
& Image Editor \\

\method{Edit Away}\pub{CVPR'25}~\cite{wang2025edit}
& White-box & Adaptive
& Laboratory
& Image Editor \\

\rowcolor{myrowcolor}
\method{Pixel Is Not a Barrier}\pub{AAAI'25}~\cite{shih2025pixel}
& Gray-box & Routine
& Laboratory
& Image Editor \\

\method{PSFD}\pub{ICME'25}~\cite{zeng2025psfd}
& Gray-box & Static
& Laboratory
& Image Editor \\

\rowcolor{myrowcolor}
\method{Semantic Mismatch}\pub{arXiv'25}~\cite{dong2025semantic}
& Black-box & Static
& External
& Image Editor \\

\method{TarPro}\pub{AAAI'26}~\cite{shen2026tarpro}
& White-box & Adaptive
& External
& Image Editor \\

\rowcolor{myrowcolor}
\method{Universal Image Immunization}\pub{arXiv'26}~\cite{lee2026universal}
& Black-box & Adaptive
& External
& Image Editor \\

\method{BlurGuard}\pub{NeurIPS'25}~\cite{kim2026blurguard}
& Black-box & Adaptive
& Laboratory
& Image Editor \\

\rowcolor{myrowcolor}
\method{Transferable Defense}\pub{TPAMI'26}~\cite{zhang2026towards}
& Gray-box & Static
& Laboratory
& Image Editor \\

\method{AdvPaint}\pub{ICLR'25}~\cite{jeon2025advpaint}
& Black-box & Adaptive
& Laboratory
& Inpainter \\

\rowcolor{myrowcolor}
\method{Anti-Inpainting}\pub{arXiv'25}~\cite{guo2025anti}
& Gray-box & Routine
& Laboratory
& Inpainter \\

\method{PromptFlare}\pub{ACM MM'25}~\cite{na2025promptflare}
& Gray-box & Adaptive
& Laboratory
& Inpainter \\

\rowcolor{myrowcolor}
\method{My Face Is Mine, Not Yours}\pub{arXiv'25}~\cite{yam2025my}
& Gray-box & Routine
& Laboratory
& Face Swapper \\

\method{Beauty and the Beast}\pub{arXiv'26}~\cite{huang2026beauty}
& Gray-box & Routine
& Laboratory
& Face Swapper \\

\rowcolor{myrowcolor}
\method{Cascading Pathway Disruption}\pub{arXiv'26}~\cite{wang2026safeguarding}
& Gray-box & Routine
& External
& Face Swapper \\

\method{I2VGuard}\pub{CVPR'25}~\cite{gui2025i2vguard}
& White-box & Adaptive
& Laboratory
& Video Animator \\

\rowcolor{myrowcolor}
\method{Anti-I2V}\pub{CVPR'26}~\cite{vu2026anti}
& Black-box & Adaptive
& Laboratory
& Video Animator \\

\method{Vid-Freeze}\pub{arXiv'25}~\cite{chowdhury2025vid}
& White-box & Routine
& Laboratory
& Video Animator \\

\rowcolor{myrowcolor}
\method{DORMANT}\pub{USENIX'25}~\cite{zhou2025dormant}
& Black-box & Adaptive
& External
& Video Animator \\

\method{Silence Is Golden}\pub{CVPR'25}~\cite{gan2025silence}
& Gray-box & Adaptive
& Laboratory
& Talking-Head Generator \\

\rowcolor{myrowcolor}
\method{SyncBreaker}\pub{arXiv'26}~\cite{zhang2026syncbreaker}
& White-box & Adaptive
& Laboratory
& Talking-Head Generator \\

\hline
\multicolumn{5}{l}{\textbf{Subject Personalization Safeguards}} \\
\hdashline

\method{IDProtector}\pub{CVPR'25}~\cite{song2025idprotector}
& Black-box & Routine
& External
& ID Adapter \\

\rowcolor{myrowcolor}
\method{DLADiff}\pub{arXiv'25}~\cite{jia2025dladiff}
& Gray-box & Static
& External
& Subject Generator \\

\method{AIR}\pub{ICME'25}~\cite{lyu2025transferable}
& Black-box & Routine
& External
& Face Swapper \\

\rowcolor{myrowcolor}
\method{Targeted Ensemble Defense}\pub{InfFus'25}~\cite{hu2025targeted}
& Gray-box & Adaptive
& Laboratory
& Subject Generator \\

\method{Adapter Shield}\pub{CVPR'26}~\cite{jia2026adapter}
& White-box & Adaptive
& Laboratory
& ID Adapter \\

\thickhline
\end{tabular}
\end{table*}

\textbf{Proactive generative safeguards} protect visual assets before release so that an existing generative pipeline cannot use them reliably as source images or conditioning references. A protector applies a visually restrained transformation to an image or a small reference set, preserving its visual utility for ordinary viewing and sharing; after release, the asset may be re-encoded, resized, or deliberately purified before an image editor, identity adapter, face swapper, video animator, or talking-head generator processes it. Protection succeeds if the pipeline can no longer produce its intended edit, identity-consistent synthesis, or subject-preserving output under the evaluated release-to-generation path. Whereas unlearnable examples limit what an unauthorized training process can learn from released data, generative safeguards limit what an already available model can produce from protected content during the generation process.

Table~\ref{tab:generative_safeguards} summarizes the surveyed methods by \textbf{transferability}, \textbf{adaptability}, \textbf{deployment readiness}, and target model family. This section includes a method only when it evaluates fixed-weight generation; five methods that also evaluate training-based customization are included for their fixed-weight branches~\cite{kim2026blurguard,jeon2025advpaint,zhou2025dormant,jia2025dladiff,hu2025targeted}.

The literature develops along two directions. (1) \emph{Editing immunization} protects an individual released image against direct semantic, identity, or motion manipulation. (2) \emph{Subject personalization safeguards} protect an identity or subject that a fixed-weight pipeline derives from one or more reference images during generation. Face manipulation lies at their boundary: direct modification of a particular image is instance-level editing, whereas extraction of a reusable identity representation is subject personalization.

\subsection{Editing Immunization}

Editing immunization modifies a released image so that a fixed-weight generator fails to perform the requested manipulation. Its \textbf{transferability} evidence spans direct optimization against a known target, transfer within a public diffusion or face-processing lineage, and transfer from surrogates to unseen targets. Its \textbf{adaptability} evidence must be read separately: model, checkpoint, or editor changes establish transferability, whereas JPEG compression, resizing, and re-encoding establish routine evidence, and protection-aware purification, removal, or retraining establishes adaptive evidence. Its \textbf{deployment readiness} remains largely at the laboratory level; several methods add human perceptual studies, and only a small subset tests external systems. None provides sustained operational evidence.

\noindent\textbf{From fixed editing to controllable manipulation.}
\method{PhotoGuard}~\cite{salman2023raising} establishes image immunization against a known editor under a fixed evaluation pipeline. Later work broadens the edited content and the controls available to a user: \method{EditShield}~\cite{chen2024editshield} studies instruction-guided editing, \method{DiffusionGuard}~\cite{choi2025diffusionguard} and \method{AdvPaint}~\cite{jeon2025advpaint} address masked or inpainting-based manipulation, and \method{Anti-Inpainting}~\cite{guo2025anti} evaluates changes in masks, prompts, seeds, and routine image transformations. \method{PromptFlare}~\cite{na2025promptflare} instead exploits cross-attention to reduce dependence on a prompt specified during protection construction. These studies broaden the generation controls considered, but variation in a mask, prompt, or seed does not by itself constitute either cross-target transfer or an adaptive countermeasure.

\noindent\textbf{Transfer across generation pipelines.}
The main transferability question is whether protection constructed on one pipeline survives when the eventual target differs. \method{PCA}~\cite{guo2025gray}, \method{Pixel Is Not a Barrier}~\cite{shih2025pixel}, and \method{PSFD}~\cite{zeng2025psfd} evaluate transfer across related editors, checkpoints, or diffusion components, supporting gray-box evidence when the source and target retain a known lineage. \method{Semantic Mismatch}~\cite{dong2025semantic}, \method{Universal Image Immunization}~\cite{lee2026universal}, \method{BlurGuard}~\cite{kim2026blurguard}, \method{AdvPaint}~\cite{jeon2025advpaint}, and \method{Anti-I2V}~\cite{vu2026anti} report direct transfer to architecturally distinct targets. \method{Transferable Defense}~\cite{zhang2026towards} also studies source-to-target transfer within a related editor ecosystem. Across these methods, a target-model change belongs to transferability even when the original paper describes it as robustness to \emph{shift}.

\noindent\textbf{Post-release transformations and informed countermeasures.}
Routine transformations and adaptive removal impose different evidentiary burdens. \method{EditShield}, \method{Pixel Is Not a Barrier}, and \method{Anti-Inpainting} evaluate protection-agnostic operations such as compression, resizing, cropping, or quantization~\cite{chen2024editshield,shih2025pixel,guo2025anti}. By contrast, \method{PCA}, \method{DiffusionGuard}, \method{DiffVax}, \method{DCT-Shield}, \method{DIA}, \method{Edit Away}, \method{TarPro}, \method{BlurGuard}, and \method{PromptFlare} explicitly test purification, restoration, or signal removal intended to weaken the protection~\cite{guo2025gray,choi2025diffusionguard,ozden2026diffvax,bala2025dct,hong2025dia,wang2025edit,shen2026tarpro,kim2026blurguard,na2025promptflare}. Frequency-domain objectives do not automatically imply post-release robustness: \method{DCT-Shield} evaluates adaptive purification and routine transformations, whereas \method{PSFD}'s audited evidence supports cross-pipeline transfer but not a qualifying post-release countermeasure~\cite{bala2025dct,zeng2025psfd}. An Adaptive label therefore records that an informed countermeasure was evaluated; it does not assert that protection remained equally strong after that countermeasure.

\noindent\textbf{Scalability and controlled failure.}
Per-image optimization can make protection expensive to apply at release time. \method{Distraction}~\cite{lo2024distraction} reduces the memory cost of image-specific optimization, \method{DiffVax}~\cite{ozden2026diffvax} amortizes protection across images, and \method{Universal Image Immunization}~\cite{lee2026universal} aims to broaden the reuse of a single protection pattern. A complementary line controls how generation fails. \method{TarPro}~\cite{shen2026tarpro} steers an edit toward a selected protection outcome, \method{DeContext}~\cite{shen2025decontext} disrupts contextual dependencies used by diffusion-transformer editors, and \method{DIA}~\cite{hong2025dia} targets inversion-based editing. These design choices improve construction cost or failure control, but neither property alone raises deployment readiness without external evidence.

\noindent\textbf{Identity editing, video, and talking-head generation.}
Editing immunization now extends beyond static semantic editing. Facial safeguards include \method{Edit Away}~\cite{wang2025edit}, \method{My Face Is Mine, Not Yours}~\cite{yam2025my}, \method{Beauty and the Beast}~\cite{huang2026beauty}, and \method{Cascading Pathway Disruption}~\cite{wang2026safeguarding}; we place them here when the evaluated misuse directly modifies a supplied face image. For image-to-video generation, \method{I2VGuard}~\cite{gui2025i2vguard} targets image-conditioned video diffusion, \method{Anti-I2V}~\cite{vu2026anti} combines color- and frequency-domain objectives, and \method{Vid-Freeze}~\cite{chowdhury2025vid} disrupts temporal evolution through the protected image. \method{DORMANT}~\cite{zhou2025dormant} covers pose-driven human animation and tests six commercial animation services, providing external evidence for those services rather than evidence of sustained deployment. \method{Silence Is Golden}~\cite{gan2025silence} and \method{SyncBreaker}~\cite{zhang2026syncbreaker} extend protection to audio-conditioned talking-head generation. Human evaluations reported by several methods establish external perceptual or utility evidence, but they likewise do not establish operational use.

\subsection{Subject Personalization Safeguards}

Within this chapter, subject personalization safeguards protect a reusable identity or subject representation that a fixed-weight generation pipeline derives from reference images. The protected reference remains recognizable to people, but an identity adapter, zero-shot subject generator, or face-swapping pipeline should produce identity-inconsistent or otherwise unusable results. In this defining branch, the protected images do not update model parameters; training-based personalization remains within UE unless the same paper also evaluates fixed-weight generation. Their \textbf{transferability} evidence ranges from target-specific construction to transfer across related adapters and independent external systems. Their \textbf{adaptability} ranges from fixed evaluation to routine transformations and informed purification or authentication bypass. Their \textbf{deployment readiness} is limited to specified services or human studies, and no method demonstrates operational deployment.

\noindent\textbf{Encoder- and adapter-based identity conditioning.}
\method{IDProtector}~\cite{song2025idprotector} uses a learned protection encoder and evaluates transfer to several identity-conditioning pipelines, including specified closed services. These tests support black-box and external evidence within the evaluated service set, not a claim about all commercial generators. \method{Adapter Shield}~\cite{jia2026adapter} adds a paired authentication mechanism so authorized generation can recover while unauthorized use fails. Its random-password bypass experiment provides limited adaptive evidence, but the authentication design is evaluated as a local prototype and therefore remains at the laboratory level.

\noindent\textbf{Extensions to training-based customization.}
\method{DLADiff}~\cite{jia2025dladiff} evaluates fixed-weight FaceID/Instance-ID generation and further extends protection to DreamBooth/LoRA customization. It also reports transfer across related Stable Diffusion settings and a 10-volunteer mean-opinion-score study. \method{Targeted Ensemble Defense}~\cite{hu2025targeted} similarly evaluates fixed-weight IP-Adapter Plus and PhotoMaker generation alongside DreamBooth/LoRA customization. Its transfer experiments remain within a known Stable Diffusion/SDXL lineage, and its adversarial-purification experiments show that protection degrades under purification. These studies demonstrate that safeguards within the scope of this section can also extend to training-based customization.

\noindent\textbf{Face-swapping boundary.}
\method{AIR}~\cite{lyu2025transferable} constructs protection with a surrogate face-recognition representation and evaluates direct transfer to independent face-swapping targets. An AWS Rekognition measurement of the swapped outputs and a human perceptual study provide two forms of external evidence; the former is an evaluation service rather than an external face-swapping target. In contrast, \method{My Face Is Mine, Not Yours}, \method{Beauty and the Beast}, and \method{Cascading Pathway Disruption} are grouped under editing immunization because their primary evaluated pipeline directly modifies a supplied image~\cite{yam2025my,huang2026beauty,wang2026safeguarding}. The distinction is therefore determined by the evaluated misuse pipeline: reusable identity extraction is subject personalization, whereas direct image modification is editing immunization.

\subsection{Discussion}

\noindent\textbf{Countermeasures.}
Generative safeguards inherit a first-mover disadvantage: the protector commits a signal before release, while a later user can transform the asset or change the generation pipeline. Protection-agnostic compression and resizing constitute routine pressure. Purification, restoration, or learned removal can deliberately attenuate the signal before generation or customization and therefore constitute adaptive pressure~\cite{cao2023impressevaluating,zhao2024can,zhao2026purify}. Changing the target editor, adapter, or backbone instead tests transferability unless it is coupled to a protection-aware operation. A defense-aware fine-tuning experiment on protected inputs, as evaluated by \method{DORMANT}~\cite{zhou2025dormant}, is a cross-stage adaptive countermeasure for that branch; the mere presence of a parallel DreamBooth/LoRA branch in another hybrid study is not, by itself, adaptive evidence.

\noindent\textbf{Open problems and future directions.}
Three gaps follow directly from the evidence in Table~\ref{tab:generative_safeguards}. First, transfer studies should report the source--target relationship explicitly, distinguishing shared checkpoint or component lineage from direct transfer to architecturally distinct or independently operated targets. Second, adaptability studies should separate routine sharing operations from informed removal, and should report the residual protection--utility trade-off rather than treating the presence of a purification experiment as proof of resistance. Third, deployment evaluation must move beyond laboratory pipelines, one-time human studies, and a small number of commercial-service tests. Only three methods report external evidence beyond human studies: \method{DORMANT} and \method{IDProtector} test specified commercial services, and \method{AIR} uses a third-party recognition service to score generated outputs; none provides sustained operational evidence. Progress therefore requires evaluations that jointly measure protection strength, visual utility, computational cost, release-platform transformations, service updates, and long-term availability without inflating prototype integration into deployment.

\section{Adversarial CAPTCHAs}
\label{sec:captcha}

\begin{table*}[!htbp]
\centering
\caption{Taxonomy of adversarial CAPTCHA methods for visual human verification.}
\label{tab:captcha_table}
\vspace{-5pt}
\tiny
\setlength{\tabcolsep}{8pt}
\renewcommand{\arraystretch}{1.0}

\begin{tabular}{l|cll|c}
\thickhline
\rowcolor{myheader}
\textbf{Paper}
& \textbf{$L_1$-Transferability}
& \multicolumn{1}{c}{\textbf{$L_2$-Adaptability}}
& \multicolumn{1}{c}{\textbf{$L_3$-Deployment Readiness}}
& \textbf{Target Model} \\
\hline

\multicolumn{5}{l}{\textbf{Character-Based CAPTCHAs}} \\
\hdashline

\rowcolor{myrowcolor}
\method{rCAPTCHA}\pub{TMM'20}~\cite{zhang2020robust}
& Black-box & Adaptive
& External
& CNN \\

\method{Spatial Smoothing}\pub{GLOBECOM'21}~\cite{matsuura2021adversarial}
& White-box & Routine
& Laboratory
& CNN \\

\rowcolor{myrowcolor}
\method{MBAGP}\pub{Electronics'21}~\cite{dankwa2021securing}
& White-box & Routine
& Laboratory
& CNN \\

\method{AECAPTCHA}\pub{JPCS'21}~\cite{wang2021text}
& White-box & Static
& Laboratory
& CNN \\

\rowcolor{myrowcolor}
\method{aCAPTCHA}\pub{TCYB'22}~\cite{shi2022adversarial}
& Gray-box & Routine
& Laboratory
& CNN \\

\method{RTC}\pub{BigData'22}~\cite{shao2022robust}
& Gray-box & Adaptive
& External
& CNN, OCR \\

\rowcolor{myrowcolor}
\method{ACG}\pub{TDSC'26}~\cite{11288041}
& Gray-box & Routine
& Laboratory
& CNN \\

\hline

\multicolumn{5}{l}{\textbf{Image-Based CAPTCHAs}} \\
\hdashline

\rowcolor{myrowcolor}
\method{DeepCAPTCHA}\pub{TIFS'17}~\cite{osadchy2017no}
& White-box & Adaptive
& External
& CNN \\

\method{Robust CAPTCHA Generator}\pub{ICAICTA'20}~\cite{ardhita2020robust}
& White-box & Routine
& Laboratory
& CNN \\

\rowcolor{myrowcolor}
\method{Capture-the-bot}\pub{IEEE Intell.\ Syst.'20}~\cite{hitaj2020capture}
& Gray-box & Static
& External
& CNN \\

\method{Diff-CAPTCHA}\pub{arXiv'23}~\cite{jiang2023diff}
& White-box & Static
& External
& CNN \\

\rowcolor{myrowcolor}
\method{DAC}\pub{TDSC'25}~\cite{du2025defensive}
& Gray-box & Adaptive
& Laboratory
& CNN \\

\hline

\multicolumn{5}{l}{\textbf{Reasoning-Based CAPTCHAs}} \\
\hdashline

\rowcolor{myrowcolor}
\method{TICS}\pub{Vis.\ Comput.'22}~\cite{jia2022tics}
& White-box & Static
& Laboratory
& CNN \\

\method{zxCAPTCHA}\pub{KST'23}~\cite{trong2023new,dinh2023zxcaptcha}
& Gray-box & Routine
& External
& CNN \\

\rowcolor{myrowcolor}
\method{IllusionCAPTCHA}\pub{WWW'25}~\cite{ding2025illusioncaptcha}
& Black-box & Static
& External
& MLLM \\

\method{Next-Gen CAPTCHAs}\pub{arXiv'26}~\cite{liu2026next}
& Black-box & Adaptive
& External
& MLLM, GUI Agent \\

\thickhline
\end{tabular}

\end{table*}

\textbf{Adversarial CAPTCHAs} apply adversarial learning to human interaction proofs that distinguish legitimate users from automated requests~\cite{lillibridge2001method,naor1996verification,von2003captcha,von2004telling}. Whereas generative safeguards protect already-released content from misuse by deployed models, adversarial CAPTCHAs guard the access gate itself: automated systems must pass the platform's verification before they can reach the visual content that the preceding stages protect. As solvers have developed from rule-based recognition to machine learning and foundation models, adversarial CAPTCHA research has expanded beyond character transcription to semantic image understanding and multimodal reasoning~\cite{chellapilla2004using,yan2008low,bursztein2011text,gao2013robustness}.

Table~\ref{tab:captcha_table} summarizes the surveyed methods by \textbf{transferability}, \textbf{adaptability}, \textbf{deployment readiness}, and target model families.

The development of adversarial CAPTCHAs can be organized into three directions. (1) \emph{Character-based CAPTCHAs} strengthen text verification against automated recognition. (2) \emph{Image-based CAPTCHAs} extend protection to semantic object recognition through adversarial and generative designs. (3) \emph{Reasoning-based CAPTCHAs} use semantic, spatial, or logical tasks to challenge multimodal models and agents.

\subsection{Character-Based CAPTCHAs}

Character-based CAPTCHAs protect text verification against automated transcription. Their \textbf{transferability} evidence spans optimization against known recognizers, transfer across solver architectures, and evaluation with third-party OCR software. Their \textbf{adaptability} ranges from fixed-pipeline evaluation to resistance against preprocessing and solver adaptation. Their \textbf{deployment readiness} remains divided between laboratory studies and external evaluation, with no sustained operational evidence in the surveyed methods.

\noindent\textbf{Baseline adversarial perturbation.}
\method{AECAPTCHA}~\cite{wang2021text} applies gradient-based perturbations against a known CNN-based character recognizer without evaluating preprocessing or solver adaptation, establishing model-directed perturbation as an early protection strategy.

\noindent\textbf{Robustness to preprocessing and solver variation.}
\method{Spatial Smoothing}~\cite{matsuura2021adversarial} targets a known CNN-based character recognizer and is designed to resist spatial smoothing. \method{rCAPTCHA}~\cite{zhang2020robust} evaluates transfer to held-out CNN-based recognizers through a segmentation--recognition pipeline. \method{aCAPTCHA}~\cite{shi2022adversarial} extends this evidence across different solver architectures and standard preprocessing filters, while \method{ACG}~\cite{11288041} evaluates transfer across multiple character recognizers and resistance to preprocessing. \method{MBAGP}~\cite{dankwa2021securing} evaluates preprocessing resistance with author-controlled CNN-based recognizers. Together, these studies move evaluation beyond a single fixed recognizer.

\noindent\textbf{Adaptive and external evaluation.}
\method{RTC}~\cite{shao2022robust} evaluates a heterogeneous set of character-recognition models, spanning shallow classifiers, neural networks, and OCR models, together with preprocessing, solver adaptation, and human usability. It therefore provides both adaptive and external evidence. Across the character-based literature, however, external evaluation does not yet amount to sustained operation in a deployed verification service.

\subsection{Image-Based CAPTCHAs}

Image-based CAPTCHAs shift the protected task from character transcription to semantic image recognition. Their \textbf{transferability} evidence ranges from known-model optimization to transfer across visual recognizers. Their \textbf{adaptability} includes fixed-pipeline evaluation, routine transformations, and informed defenses. Their \textbf{deployment readiness} is supported by both laboratory experiments and human-facing external evaluation, but not by operational deployment.

\noindent\textbf{Perturbation-based protection.}
\method{DeepCAPTCHA}~\cite{osadchy2017no} optimizes adversarial noise against a known CNN-based image recognizer and evaluates resistance to noise removal. \method{Robust CAPTCHA Generator}~\cite{ardhita2020robust} optimizes against a known image recognizer while accounting for geometric and photometric transformations. \method{Capture-the-bot}~\cite{hitaj2020capture} evaluates localized adversarial patterns across several CNN-based image recognizers without a targeted solver countermeasure. These methods broaden perturbation-based evaluation from a fixed recognizer to transformations, signal removal, and cross-architecture transfer.

\noindent\textbf{Generative protection and stronger solver tests.}
\method{Diff-CAPTCHA}~\cite{jiang2023diff} incorporates adversarial protection into diffusion-based generation and evaluates the resulting challenges with author-controlled CNN-based image recognizers. \method{DAC}~\cite{du2025defensive} evaluates its challenges across multiple CNN-based image recognizers and several defensive preprocessing methods. These studies distinguish the mechanism used to generate a challenge from the system used to solve it.

\noindent\textbf{Human-facing evaluation.}
Human-facing studies accompany the machine-side evaluations of \method{DeepCAPTCHA}, \method{Capture-the-bot}, and \method{Diff-CAPTCHA}~\cite{osadchy2017no,hitaj2020capture,jiang2023diff}. These studies provide external evidence through usability or perceptual-quality assessment, but they do not provide operational deployment evidence.

\subsection{Reasoning-Based CAPTCHAs}

Reasoning-based CAPTCHAs move the solver bottleneck from recognition alone toward semantic, spatial, or logical interpretation. The surveyed methods include designs evaluated with CNN-based image recognizers and later challenges evaluated with proprietary vision--language systems and GUI-agent systems. Their \textbf{transferability} therefore ranges from white-box to black-box evaluation, their \textbf{adaptability} ranges from static tests to informed solving strategies, and their \textbf{deployment readiness} remains laboratory or external rather than operational.

\noindent\textbf{Recognition-oriented designs.}
\method{TICS}~\cite{jia2022tics} is evaluated with an author-trained CNN-based image recognizer under a fixed laboratory setting. \method{zxCAPTCHA}~\cite{trong2023new,dinh2023zxcaptcha} extends evaluation across several CNN-based image recognizers, preprocessing defenses, and human users. These studies retain recognition-based solvers while increasing the semantic structure of the challenge.

\noindent\textbf{Vision--language and agentic solvers.}
\method{IllusionCAPTCHA}~\cite{ding2025illusioncaptcha} uses visual illusions and evaluates stock proprietary vision--language systems together with human users. \method{Next-Gen CAPTCHAs}~\cite{liu2026next} evaluates commercial vision--language systems and GUI-agent systems on spatial and multi-attribute tasks. It also tests an informed, family-aware hint strategy, providing adaptive evidence beyond fixed-model evaluation.

\noindent\textbf{From recognition failure to solving cost.}
Reasoning-oriented evaluation considers human performance alongside the computational effort required by vision--language and GUI-agent systems~\cite{trong2023new,dinh2023zxcaptcha,ding2025illusioncaptcha,liu2026next}. This extends evaluation beyond whether one fixed solver succeeds, while retaining the central requirement that legitimate users can complete the challenge.

\subsection{Discussion}

\noindent\textbf{Countermeasures.}
The surveyed evidence identifies three pressures on adversarial CAPTCHA protection. Routine preprocessing can weaken perturbations before recognition, restoration or defensive preprocessing can target the protective signal, and adapted solvers can retrain or change their solving strategy~\cite{matsuura2021adversarial,shi2022adversarial,ardhita2020robust,du2025defensive,shao2022robust,liu2026next}. Results obtained against a fixed recognizer therefore do not establish robustness against an informed solver.

\noindent\textbf{Open problems and future directions.}
Three gaps remain. Transferability evaluation should distinguish known targets, cross-architecture transfer, and genuinely external solvers more consistently. Adaptability evaluation should test both routine preprocessing and informed solver adjustment, especially as vision--language and GUI-agent systems replace conventional recognizers. Deployment evidence should also move beyond laboratory and one-time user studies toward long-term operation, while reporting human usability together with solver resistance. These requirements follow the same three-axis framework used throughout this survey and avoid treating success against a single solver as evidence of general protection.

\section{Adversarial Provenance and Accountability}
\label{sec:provenance}
\begin{table*}[!htbp]
\centering
\caption{Taxonomy of adversarial provenance and accountability methods.}
\label{tab:provenance-accountability}
\vspace{-5pt}
\tiny
\setlength{\tabcolsep}{4pt}
\renewcommand{\arraystretch}{1.1}

\begin{tabular}{l|cll|c}

\thickhline
\rowcolor{myheader}

\textbf{Paper}
& \textbf{$L_1$-Transferability}
& \multicolumn{1}{c}{\textbf{$L_2$-Adaptability}}
& \multicolumn{1}{c}{\textbf{$L_3$-Deployment Readiness}}
& \textbf{Target Model} \\

\hline

\multicolumn{5}{l}{\textbf{Training Asset Tracing}} \\
\hdashline

\rowcolor{myrowcolor}
\method{Radioactive Data}\pub{ICML'20}~\cite{sablayrolles2020radioactivedata}
& Gray-box & Media Transformations
& Benchmark-scale
& Vision DNN \\

\method{Dataset Inference}\pub{arXiv'21}~\cite{maini2021datasetinference}
& Black-box Query & Model Adaptation
& Benchmark-scale
& Vision DNN \\

\rowcolor{myrowcolor}
\method{DVBW}\pub{TIFS'23}~\cite{li2023blackbox}
& Gray-box & No Adaptive Eval.
& Benchmark-scale
& Vision DNN \\

\method{SSCL-BW}\pub{arXiv'25}~\cite{wang2025ssclbwsample}
& Gray-box & Evidence Manipulation
& Benchmark-scale
& Vision DNN \\

\rowcolor{myrowcolor}
\method{X-Mark}\pub{arXiv'26}~\cite{kulkarni2026xmarksaliency}
& Gray-box & Evidence Manipulation
& Benchmark-scale
& Vision DNN \\

\hline

\multicolumn{5}{l}{\textbf{Model Ownership Verification}} \\
\hdashline

\rowcolor{myrowcolor}
\method{DNN Watermark}\pub{AsiaCCS'18}~\cite{zhang2018protectingintellectual}
& Gray-box & No Adaptive Eval.
& Lab-only
& Vision DNN \\

\method{IPN Watermark}\pub{AAAI'20}~\cite{zhang2020modelwatermarking}
& Gray-box & Model Adaptation
& Benchmark-scale
& Vision DNN \\

\rowcolor{myrowcolor}
\method{Deep Watermarking}\pub{TPAMI'21}~\cite{zhang2021deepmodel}
& Gray-box & Model Adaptation
& Benchmark-scale
& Vision DNN \\

\method{Wide-Flat GAN WM}\pub{TIFS'24}~\cite{fei2023wideflat}
& Gray-box & Model Adaptation
& Benchmark-scale
& Image/Video Generator \\

\rowcolor{myrowcolor}
\method{Free Fine-tuning WM}\pub{ACM MM'23}~\cite{wang2023freefine}
& Gray-box & Evidence Manipulation
& Benchmark-scale
& Vision DNN \\

\method{CycleGAN Watermark}\pub{TDSC'24}~\cite{lin2024cycleganwatermarking}
& Gray-box & Model Adaptation
& Lab-only
& Image/Video Generator \\

\rowcolor{myrowcolor}
\method{PlugMark}\pub{ICCV'25}~\cite{chen2025plugmarkplug}
& Gray-box & Model Adaptation
& Benchmark-scale
& Image/Video Generator \\

\method{VLA-Mark}\pub{EMNLP'25}~\cite{liu2025vlamarkcross}
& Gray-box & Evidence Manipulation
& Lab-only
& VLM/MLLM \\

\rowcolor{myrowcolor}
\method{SWAP}\pub{arXiv'25}~\cite{yang2025swapcopyrightauditing}
& Gray-box & No Adaptive Eval.
& Benchmark-scale
& VLM/MLLM \\

\method{Cert-LAS}\pub{arXiv'26}~\cite{qi2026certlascertified}
& Gray-box & Evidence Manipulation
& Benchmark-scale
& Image/Video Generator \\

\rowcolor{myrowcolor}
\method{VLA Backdoor Ownership}\pub{arXiv'26}~\cite{sun2026backdoorbasedownership}
& Gray-box & Model Adaptation
& Benchmark-scale
& VLM/MLLM \\

\method{LoRA-Key}\pub{arXiv'26}~\cite{wang2026lorakeyuser}
& Gray-box & Model Adaptation
& Benchmark-scale
& Image/Video Generator \\

\rowcolor{myrowcolor}
\method{SIF}\pub{arXiv'26}~\cite{zhao2026sifsemanticallydistribution}
& Gray-box & Model Adaptation
& Benchmark-scale
& VLM/MLLM \\

\hline
\multicolumn{5}{l}{\textbf{Media Provenance Verification}} \\
\hdashline

\rowcolor{myrowcolor}
\method{ROBIN}\pub{NeurIPS'24}~\cite{huang2024robinrobust}
& Gray-box & Evidence Manipulation
& Benchmark-scale
& Image/Video Generator \\

\method{ConceptWM}\pub{arXiv'24}~\cite{lei2024watermarkingvisual}
& Gray-box & Evidence Manipulation
& Benchmark-scale
& Image/Video Generator \\

\rowcolor{myrowcolor}
\method{Traceable Adv. Examples}\pub{TCSVT'24}~\cite{li2024dualprotection}
& Content-only & Media Transformations
& Benchmark-scale
& Image/Video Generator \\

\method{Invisible Adv. WM}\pub{TOMM'24}~\cite{wang2024invisibleadversarial}
& Gray-box & No Adaptive Eval.
& Lab-only
& Vision DNN \\

\rowcolor{myrowcolor}
\method{ZoDiac}\pub{NeurIPS'24}~\cite{zhang2024attackresilient}
& Gray-box & Evidence Manipulation
& Benchmark-scale
& Image/Video Generator \\

\method{NoisePrints}\pub{arXiv'25}~\cite{goren2025noiseprintsdistortionfree}
& Gray-box & Evidence Manipulation
& Benchmark-scale
& Image/Video Generator \\

\rowcolor{myrowcolor}
\method{Video Signature}\pub{arXiv'25}~\cite{huang2025videosignatureimplicit}
& Gray-box & Evidence Manipulation
& Benchmark-scale
& Image/Video Generator \\

\method{BitMark}\pub{NeurIPS'26}~\cite{kerner2025bitmarkwatermarkingbitwise}
& Gray-box & Evidence Manipulation
& Benchmark-scale
& Image/Video Generator \\

\rowcolor{myrowcolor}
\method{Info-Theoretic AIGI Detector}\pub{arXiv'25}~\cite{zhang2025adversariallyrobustai}
& Content-only & Evidence Manipulation
& Benchmark-scale
& Image/Video Generator \\

\method{CSGuard}\pub{arXiv'26}~\cite{lai2026csguardforgeryresistant}
& Gray-box & Evidence Manipulation
& Lab-only
& Image/Video Generator \\

\rowcolor{myrowcolor}
\method{RWP}\pub{Neural Networks'26}~\cite{liu2026rwprobust}
& Gray-box & Evidence Manipulation
& Benchmark-scale
& Image/Video Generator \\

\method{AEON}\pub{WACV'26}~\cite{muneer2026aeonadaptive}
& Gray-box & Evidence Manipulation
& Lab-only
& Image/Video Generator \\

\rowcolor{myrowcolor}
\method{Dual-Guard}\pub{arXiv'26}~\cite{xie2026dualguard}
& Content-only & Evidence Manipulation
& Benchmark-scale
& Image/Video Generator \\

\method{Robust Content WM}\pub{arXiv'26}~\cite{zhu2026robustcontentwatermarking}
& Gray-box & Evidence Manipulation
& Benchmark-scale
& Image/Video Generator \\

\hline
\multicolumn{5}{l}{\textbf{Evidence Manipulation Attacks}} \\
\hdashline

\rowcolor{myrowcolor}
\method{Watermark Radioactivity Eval.}\pub{ICLR'25 Workshop}~\cite{dubiski2025arewatermarks}
& Gray-box & Model Adaptation
& Lab-only
& Image/Video Generator \\

\method{Dataset Copyright Evasion}\pub{TIFS'25}~\cite{gao2025datasetcopyrightevasion}
& Gray-box & Evidence Manipulation
& Benchmark-scale
& Image/Video Generator \\

\rowcolor{myrowcolor}
\method{FT-Traceability Benchmark}\pub{CVPR'26}~\cite{wang2025evaluatingdatasetwatermarking}
& Gray-box & Evidence Manipulation
& Benchmark-scale
& Image/Video Generator \\

\method{MarkSweep}\pub{ICASSP'26}~\cite{cao2026marksweepnobox}
& Gray-box & Evidence Manipulation
& Benchmark-scale
& Image/Video Generator \\

\rowcolor{myrowcolor}
\method{Forensic-Stealth WM Removal}\pub{arXiv'26}~\cite{goonatilake2026removingwatermarkis}
& Gray-box & Evidence Manipulation
& Benchmark-scale
& Image/Video Generator \\

\method{Fragile Reconstruction}\pub{arXiv'26}~\cite{jiang2026fragilereconstructionadversarial}
& Black-box Query & Evidence Manipulation
& Lab-only
& Image/Video Generator \\

\rowcolor{myrowcolor}
\method{RAVEN}\pub{arXiv'26}~\cite{shamshad2026ravenerasinginvisible}
& Gray-box & Evidence Manipulation
& Benchmark-scale
& Image/Video Generator \\

\method{Frequency-Domain WM Attack}\pub{arXiv'26}~\cite{wang2026breakingwatermarksfrequency}
& Gray-box & Evidence Manipulation
& Benchmark-scale
& Image/Video Generator \\

\thickhline
\end{tabular}
\end{table*}

When privacy filters, unlearnable data, or other prevention-oriented defenses fail, visual assets may already have entered training corpora, model services, or public content streams. \textbf{Adversarial provenance and accountability} address this downstream stage: they ask how a data owner, model developer, platform, or user can later prove data use, model copying, content origin, or media manipulation. We focus on evidence that is adversarially constructed, verified against an active counterparty, or stress-tested by attacks, including dataset traces, ownership triggers, diffusion watermarks, and watermark red teaming~\cite{sablayrolles2020radioactivedata,zhang2018protectingintellectual,huang2024robinrobust}. The goal is not to promise an indestructible mark, but to make misuse harder to deny under realistic disputes.

Table~\ref{tab:provenance-accountability} summarizes representative work through three layers. \textbf{L1 Transferability} asks whether verification can move from internal access to owner-keyed, black-box, or content-only disputes. \textbf{L2 Adaptability} asks whether evidence remains meaningful under model adaptation, removal, forgery, or semantic manipulation. \textbf{L3 Deployment Readiness} asks whether the evidence stays in benchmarks or approaches service, platform, and arbitration workflows. We organize this section around three application scenarios: \emph{(1) adversarial training-asset tracing}, which audits whether visual data were used for training; \emph{(2) adversarial model ownership verification}, which supports claims over copied or adapted model components; and \emph{(3) attack-resilient generated-media provenance}, which traces AIGC images and videos after circulation and manipulation~\cite{maini2021datasetinference,zhang2020modelwatermarking,zhang2024attackresilient}.

\subsection{Adversarial Training-Asset Tracing}

Adversarial training-asset tracing is the post-hoc counterpart of data protection: instead of stopping unauthorized learning, it tries to show that a released visual dataset, annotation set, or sensitive collection contributed to a trained model~\cite{sablayrolles2020radioactivedata,maini2021datasetinference}. The scenario has become practical because web-scale vision and vision-language pipelines mix scraped images, licensed datasets, synthetic samples, and private collections before an external auditor can inspect the model. The application story is ordered by how much control the owner has over the data path: licensed release, web-scale scraping, and mixed-corpus attribution.

\noindent\textbf{Licensed dataset release and ownership disputes.}
When a dataset owner intentionally releases or licenses visual data, the key application question is how to keep a later ownership claim possible. This is the cleanest tracing setting because the owner can prepare the release. \method{Radioactive Data}~\cite{sablayrolles2020radioactivedata} shows that small optimized changes to released images can leave a statistical trace in later models. Dataset watermarking then makes the claim more operational: \method{DVBW}~\cite{li2023blackbox} uses backdoor-style marks so an owner can query a suspected model, while clean-label variants such as \method{SSCL-BW}~\cite{wang2025ssclbwsample} and \method{X-Mark}~\cite{kulkarni2026xmarksaliency} study stronger evidence under sample modification or medical-domain constraints.

\noindent\textbf{Web-scale scraping and post-hoc audits.}
For data already circulating on the web, the owner may have no chance to premark the release. \method{Dataset Inference}~\cite{maini2021datasetinference} instead tests whether a model's black-box behavior reflects a particular dataset, shifting the dispute from a secret mark to statistical evidence. This branch fits post-hoc audits of scraped training data, but it also makes the claim more probabilistic and sensitive to model adaptation.

\noindent\textbf{Mixed-corpus attribution under weak evidence.}
The least controlled setting is a real training pipeline, where the suspected data may be mixed with public pretraining data, duplicates, synthetic samples, and later fine-tuning sets. Data-use claims must therefore account for deduplication and copyright-evasion strategies~\cite{gao2025datasetcopyrightevasion,wang2025evaluatingdatasetwatermarking}. Thus the main deployment challenge is not merely detecting use, but distinguishing misuse from common pretraining, duplicated web images, and distributional similarity.

\subsection{Adversarial Model Ownership Verification}

Adversarial model ownership verification asks whether a developer can prove that a deployed model, API, adapter, or derivative service came from their original visual model. This scenario appears when checkpoints are copied, APIs are extracted, models are fine-tuned, or components are repackaged in model hubs and service markets. The disputed object has expanded from a single classifier to image-processing networks, GANs, diffusion models, VLMs, VLAs, prompts, and LoRA modules~\cite{zhang2020modelwatermarking,fei2023wideflat,liu2025vlamarkcross,wang2026lorakeyuser}. The application story follows where ownership disputes now occur, moving from whole-model copying to derivative services and then to component reuse in VLM/VLA and adapter ecosystems.

\noindent\textbf{Model marketplace and API-copying disputes.}
In model marketplaces or hosted APIs, the suspicious artifact may be a copied checkpoint or extracted service rather than a file the owner can inspect. Early model watermarks turn ownership into a behavioral test: the DNN watermarking framework~\cite{zhang2018protectingintellectual} uses trigger-like inputs for remote claims, while \method{IPN Watermarking}~\cite{zhang2020modelwatermarking} and deep output-barrier watermarking~\cite{zhang2021deepmodel} make stolen image-processing networks inherit owner-visible output patterns. Free fine-tuning watermarking later targets disputes where the accused model has been modified or overwritten before verification~\cite{wang2023freefine}.

\noindent\textbf{Derivative generative services.}
The next setting is a GAN or diffusion service, where copying is harder to define because the suspect may be a fine-tuned generator, a merged checkpoint, or a model that only preserves part of the original behavior. \method{Wide-flat GAN watermarking}~\cite{fei2023wideflat} uses adversarial parameter noise to keep ownership evidence stable under model updates, CycleGAN watermarking studies adapted image generators~\cite{lin2024cycleganwatermarking}, and diffusion ownership work moves toward derivative verification and certified removal resistance~\cite{chen2025plugmarkplug,qi2026certlascertified}. Red-team evaluation further shows that watermark radioactivity can be weakened by model adaptation~\cite{dubiski2025arewatermarks}. In application terms, the evidence must survive the same operations that make derivative services commercially useful: fine-tuning, pruning, extraction, overwriting, and surrogate training.

\noindent\textbf{Component reuse in model ecosystems.}
The most modular setting is a modern AI service, where the disputed asset may be a reusable component rather than a whole model. \method{PlugMark}~\cite{chen2025plugmarkplug} verifies diffusion derivatives through decision-boundary zero-watermarks, while VLM/VLA and adapter work attaches ownership to cross-modal outputs, soft prompts, embodied observations, and LoRA modules~\cite{liu2025vlamarkcross,yang2025swapcopyrightauditing,sun2026backdoorbasedownership,wang2026lorakeyuser}. \method{SIF}~\cite{zhao2026sifsemanticallydistribution} further studies semantically in-distribution fingerprints for VLMs. The remaining application tension is clear: owner secrets help resist evasion, but public arbitration needs evidence that a neutral party can reproduce without leaking reusable triggers or keys.

\subsection{Attack-Resilient Generated-Media Provenance}

Attack-resilient generated-media provenance addresses the public side of accountability: after an AI-generated image or video leaves the generator, it may be edited, regenerated, screenshotted, compressed, reposted, or stripped of evidence~\cite{zhang2024attackresilient,kerner2025bitmarkwatermarkingbitwise,huang2025videosignatureimplicit}. The core question is no longer whether a clean watermark can be decoded, but whether a platform, creator, journalist, or victim can still support an origin claim after adversarial circulation. The application story follows the media lifecycle: a creator first needs a claim after circulation, an adversary may then create false attribution, and a platform finally has to review partial evidence.

\noindent\textbf{Creator-side claims after content circulation.}
Creators and generation services need provenance after media have passed through editing tools, social platforms, recompression, reposting, and adversarial watermark erasure. \method{ROBIN}~\cite{huang2024robinrobust} and \method{ZoDiac}~\cite{zhang2024attackresilient} optimize diffusion watermarks against removal and regeneration, while traceable adversarial examples and invisible adversarial watermarks turn perturbations into copyright evidence~\cite{li2024dualprotection,wang2024invisibleadversarial}. The practical principle is that provenance should be designed for the path media will actually take, not only for clean decoding immediately after generation.

\noindent\textbf{False attribution and ownership disputes.}
After circulation, the next dispute is adversarial attribution: attackers may not only remove provenance but also forge or transfer it. \method{ConceptWM}~\cite{lei2024watermarkingvisual} protects visual-concept watermarks under purification and fine-tuning, \method{NoisePrints}~\cite{goren2025noiseprintsdistortionfree} binds authorship to secret diffusion seeds, and \method{CSGuard}~\cite{lai2026csguardforgeryresistant} targets forgery-resistant watermark recovery. These methods shift the goal from benign robustness to dispute safety: a useful signal should resist both disappearance and false attribution.

\noindent\textbf{Platform moderation and forensic review.}
At the end of the lifecycle, platforms, journalists, or moderators often only possess the circulated image or video, not the original prompt, model, or owner key. \method{BitMark}~\cite{kerner2025bitmarkwatermarkingbitwise} embeds bit-level evidence in autoregressive generation, video signatures treat temporal tampering as a first-class threat~\cite{huang2025videosignatureimplicit}, and adaptive watermarking studies cumulative attacks and removal-forgery tradeoffs~\cite{muneer2026aeonadaptive,zhu2026robustcontentwatermarking}. Content-only methods such as adversarially robust AIGI detection and \method{Dual-Guard}~\cite{zhang2025adversariallyrobustai,xie2026dualguard} support moderation or forensic review without generator access. At deployment time, these signals may trigger labeling, escalation, creator notification, or human review rather than an automatic verdict, especially when watermark erasure tools are also available~\cite{liu2026rwprobust,shamshad2026ravenerasinginvisible}.

\subsection{Discussion}

\noindent\textbf{Countermeasures and limitations.}
The main limitation of adversarial provenance is that evidence becomes another attack surface. Dataset traces may be diluted by dataset mixing or evasion~\cite{gao2025datasetcopyrightevasion,wang2025evaluatingdatasetwatermarking}; model watermarks may be weakened by fine-tuning, extraction, or removal~\cite{dubiski2025arewatermarks,qi2026certlascertified}; and media watermarks can fail under no-box removal, forensic-stealth attacks, reconstruction, or frequency-domain filtering~\cite{cao2026marksweepnobox,goonatilake2026removingwatermarkis,jiang2026fragilereconstructionadversarial,wang2026breakingwatermarksfrequency}. Existing countermeasures, such as certified verification, author-proof designs, anti-forgery constraints, and robust content watermarking, protect useful parts of the chain~\cite{goren2025noiseprintsdistortionfree,lai2026csguardforgeryresistant,zhu2026robustcontentwatermarking}, but they rarely provide end-to-end accountability across data owners, model hubs, service providers, platforms, and victims. A protocol remains hard to deploy if it cannot explain who may run the test, what secrets must be revealed, how uncertainty is reported, and how the accused party can contest the result.

\noindent\textbf{Open problems and future directions.}
The useful lesson for AIGC and agent-era safety is that adversarial thinking should shape the evidence workflow, not only the watermark algorithm. Future provenance systems should connect data-use traces, model-component ownership, adapter reuse, tool calls, generated media, and platform logs into evidence chains~\cite{liu2025vlamarkcross,sun2026backdoorbasedownership,wang2026lorakeyuser,liu2026rwprobust}. This is especially relevant for native multimodal systems, world-models, vision-language-actions, and multimodal agents: a GUI agent may learn from screenshots and action logs, a robot policy may inherit visual demonstrations, and a video-generation service may reuse private adapters or scene priors. For such systems, provenance cannot treat media files as isolated artifacts; it must bind perception, decision, tool use, and generation records into an auditable workflow. Stronger \textbf{L1 Transferability} requires verification that works for auditors without exposing reusable secrets; stronger \textbf{L2 Adaptability} requires composed attacks that combine model adaptation, regeneration, false attribution, and trace tampering; stronger \textbf{L3 Deployment Readiness} requires API versioning, registries, content-platform workflows, and third-party arbitration. The long-term goal is not a permanent signal, but a calibrated body of evidence whose uncertainty remains interpretable when adversaries attack the weakest link.

\section{Conclusion}
\label{sec:conclusion}

This survey has followed a single idea across five research communities: the perturbations that expose the fragility of learned models can be turned, by the owners of visual content, into protection. Read in lifecycle order, adversarial privacy filters, unlearnable examples, generative safeguards, adversarial CAPTCHAs, and provenance mechanisms are not isolated literatures but successive answers to the same structural asymmetry, applied at the moments of sharing, release, generation, access, and dispute (Sections~\ref{sec:privacy}--\ref{sec:provenance}). The shared axes of transferability, adaptability, and deployment readiness ($L_1$--$L_3$) let their threat models and robustness claims be weighed on one scale.

\noindent\textbf{Cross-stage countermeasures.} Seen side by side, the five families face the same three counterattacks. Adversaries purify, restore, or re-encode a protected asset until a near-clean copy re-enters the pipeline; they swap the pipeline itself, changing backbones, training objectives, or customization protocols until the assumptions behind a protection no longer hold; and once a protective signal can be detected, they strip it or, in the provenance setting, forge it. Beneath all three lies an asymmetry that the protective inversion itself creates: the protector commits a transformation at release time and cannot revise it, while the adversary moves second, with the asset in hand and unlimited attempts. Static validation therefore flatters every mechanism in this survey. Robustness claims are meaningful only against informed adversaries, and they remain conditional on the capability the adversary is assumed to have.

\noindent\textbf{Cross-stage open problems.} Four problems recur across the five stages and will decide how far the paradigm carries. First, evaluation is still incommensurable in practice: each community reports robustness against its own surrogates, datasets, and attack suites; adaptive evaluation is the exception; and evidence seldom matures beyond the laboratory. Shared benchmarks that instantiate $L_1$--$L_3$ with informed attackers are the most immediate need. Second, the field should retire the promise of absolute prevention. Where removal is possible given enough capability, the honest measures of protection are the cost it imposes, the authorization it can grant or withhold, and the evidence it leaves behind; prevention up front and provenance behind it are layers of one design rather than competing goals. Third, the lifecycle view exposes a question no single community can ask: how protective signals compose. A photograph may need to defeat recognition, resist training, disrupt personalization, and still carry a verifiable mark, all within one imperceptibility budget, yet almost nothing is known about how such signals interfere or how the budget should be divided among them. Fourth, the target is moving. Every stage reports the same shift, away from fixed recognizers and editors toward multimodal models and autonomous agents that can re-perceive, reason about, and retry a protected asset; protection designed to survive one inference pass must now hold against a compositional stack.

\noindent\textbf{Outlook.} The reason to expect this paradigm to endure is the same reason adversarial examples were never engineered away: the perceptual gap between human observers and learned models is a structural property of gradient-trained systems, and every pipeline that automates the use of visual content inherits it. Whoever controls an asset before release can therefore reach into pipelines they will never see, and each new class of systems, multimodal or agentic, renews the opportunity along with the risk. The paradigm's name is meant in both of its senses: these are attacks turned to good ends, and, for as long as AI is built this way, attacks that are here for good.

\bibliographystyle{TPAMI-Reference-Format}
\bibliography{ref}

\end{document}